\documentclass[onecolumn,superscriptaddress,prd,aps,preprintnumbers,amsmath,amssymb,nofootinbib]{revtex4}
\usepackage{graphicx}
\usepackage{dcolumn}
\usepackage{bm}
\usepackage{bbm}
\usepackage[dvips]{color}

\newcommand{\beq}{\begin{equation}}
\newcommand{\eeq}{\end{equation}}
\newcommand{\bq}{\begin{equation}}
\newcommand{\eq}{\end{equation}}
\newcommand{\ba}{\begin{array}}
\newcommand{\ea}{\end{array}}
\newcommand{\beqa}{\begin{eqnarray}}
\newcommand{\eeqa}{\end{eqnarray}}

\def\[{\left[}
\def\]{\right]}
\def\({\left(}
\def\){\right)}

\evensidemargin=\oddsidemargin

\def\slashi#1{\rlap{\sl/}#1}
\def\slashii#1{\setbox0=\hbox{$#1$}             
   \dimen0=\wd0                                 
   \setbox1=\hbox{\sl/} \dimen1=\wd1            
   \ifdim\dimen0>\dimen1                        
      \rlap{\hbox to \dimen0{\hfil\sl/\hfil}}   
      #1                                        
   \else                                        
      \rlap{\hbox to \dimen1{\hfil$#1$\hfil}}   
      \hbox{\sl/}                               
   \fi}                                         %
\def\slashiii#1{\setbox0=\hbox{$#1$}#1\hskip-\wd0\hbox to\wd0{\hss\sl/\/\hss}}
\def\pslash{\not{\hbox{\kern-4pt $p$}}}
\def\qslash{\not{\hbox{\kern-4pt $q$}}}
\def\lv{\not{\hbox{\kern-4pt $L$}}}
\def\lsim{\mathrel{\raise.3ex\hbox{$<$\kern-.75em\lower1ex\hbox{$\sim$}}}}
\def\gsim{\mathrel{\raise.3ex\hbox{$>$\kern-.75em\lower1ex\hbox{$\sim$}}}}
\def\ifmath#1{\relax\ifmmode #1\else $#1$\fi}
\newcommand{\tbox}[1]{\mbox{\tiny #1}}
\newcommand{\bL}{b_{\tbox{L}}}
\newcommand{\bR}{b_{\tbox{R}}}
\newcommand{\BL}{B_{\tbox{L}}}
\newcommand{\BR}{B_{\tbox{R}}}
\newcommand{\bLb}{\overline{b}_{\tbox{L}}}
\newcommand{\bRb}{\overline{b}_{\tbox{R}}}
\newcommand{\BLb}{\overline{B}_{\tbox{L}}}
\newcommand{\BRb}{\overline{B}_{\tbox{R}}}

\newcommand{\PL}{P_{\tbox{L}}}
\newcommand{\PR}{P_{\tbox{R}}}
\newcommand{\slh}[1]{\displaystyle{\not}#1}

\newcommand{\ii}{\mathbbm{i}}

\newcommand{\pz}{\pi_{\tbox{$Z$}}}

\newcommand{\vev}[1]{\langle 0 | #1 |0\rangle}

\begin{document}

\preprint{NTLP 2008-04}
\preprint{MSUHEP-090223}

 \title{$Z\to b\bar{b}$ and Chiral Currents in Higgsless Models}

\author{Tomohiro Abe}
\affiliation{Department of Physics, Nagoya University, Nagoya 464-8602, Japan}

\author{R.\ Sekhar Chivukula}
\author{Neil D.~Christensen}
\author{Ken Hsieh}
\affiliation{Department of Physics and Astronomy, Michigan State
       University, East Lansing, MI 48824, USA}
       
\author{Shinya Matsuzaki}
\affiliation{Department of Physics and Astronomy, 
  University of North Carolina, Chapel Hill, NC 27599, USA}

\author{Elizabeth H. Simmons}
\affiliation{Department of Physics and Astronomy, Michigan State
       University, East Lansing, MI 48824, USA}

\author{Masaharu Tanabashi}  
\affiliation{Department of Physics, Nagoya University, Nagoya 464-8602, Japan}

\date{\today}

\begin{abstract}
In this note we compute the flavor-dependent chiral-logarithmic corrections to the decay
$Z \to b\bar{b}$ in the three site Higgsless model. We compute
these corrections diagrammatically in the ``gaugeless" limit in which the 
electroweak couplings vanish. We also compute the chiral-logarithmic corrections to the
decay $Z\to b\bar{b}$ using an RGE analysis in effective field theory, and show that the results 
agree.  In the process of this computation, 
we compute the form of the chiral current in the gaugeless limit 
of the three-site model, and consider
the generalization to the $N$-site case. We elucidate the
Ward-Takahashi identities which underlie the gaugeless limit calculation in the three-site model,
and describe how the result for the $Z\to b\bar{b}$ amplitude
is obtained in unitary gauge in the full theory.
We find that the phenomenological constraints on the three-site Higgsless
model arising from measurements of $Z\to b\bar{b}$ are relatively mild, requiring
only that the heavy Dirac fermion be heavier than 1 TeV or so, and are satisfied
automatically in the range of parameters allowed by other precision electroweak
data. 
\end{abstract}

\date{\today}
 
 \maketitle


\section{Introduction}

Higgsless models \cite{Csaki:2003dt} of electroweak symmetry breaking provide
effective low-energy theories of a strongly-interacting symmetry
breaking sector \cite{Weinberg:1979bn,Susskind:1978ms} which, in the
case of  ``delocalized" fermions
\cite{Cacciapaglia:2004rb,Casalbuoni:2005rs,Cacciapaglia:2005pa,Foadi:2004ps,Foadi:2005hz,Chivukula:2005bn,SekharChivukula:2005xm}, can be consistent with electroweak precision
measurements  \cite{SekharChivukula:2006cg,Abe:2008hb}. 
The three-site model \cite{SekharChivukula:2006cg}
is the minimal low-energy realization of a  
Higgsless theory.  It includes only the lightest triplet of the extra  
vector measons typically present in such theories; these are the
mesons (denoted here by ${W'}^\pm$ and  $Z'$) that are analogous to the $\rho$ mesons of QCD.
The three-site model retains sufficient complexity, however,
to incorporate interesting physics issues related to fermion masses and electroweak observables.
In particular the chiral logarithmic corrections -- the one-loop contributions which dominate in the limit
$M_{W'} \ll \Lambda$ where $M_{W'}$ are the masses of the new vector mesons 
and $\Lambda$ is the cutoff of the effective theory -- 
to the flavor-universal electroweak parameters
$\alpha S$ and $\alpha T$ 
\cite{Peskin:1990zt,Peskin:1991sw,Altarelli:1990zd,Altarelli:1991fk} 
in the three-site model were computed
in references \cite{Matsuzaki:2006wn,SekharChivukula:2007ic,Dawson:2007yk}.

In this note we compute the flavor-dependent chiral logarithmic corrections
to the process $Z \to b\bar{b}$ in the three-site model. We perform the computation diagrammatically
in the ``gaugeless"
limit \cite{Lytel:1980zh,Barbieri:1992nz,Barbieri:1992dq,Oliver:2002up}, in which the electroweak couplings vanish: here the corrections to
the $Z$-boson coupling are related to the couplings of the bottom-quark to the
neutral Nambu-Goldstone boson
present in this limit. As we illustrate, the computation of the amplitude is
complicated by the mixing of the light fermions with the heavy Dirac fermions
present in the three-site model; such issues do not arise in many other 
theories beyond the Standard Model,  
such as the MSSM or models featuring extended electroweak gauge groups  
but no new fermions.
We also compute the chiral-logarithmic corrections to the
decay $Z\to b\bar{b}$ in effective field theory and show that the results 
agree with the diagrammatic computation. 
We find that the phenomenological constraints on the three-site Higgsless
model arising from measurements of $Z\to b\bar{b}$ are relatively mild, requiring
only that the heavy Dirac fermion be heavier than 1 TeV or so, and are satisfied
automatically in the range of parameters allowed by other precision electroweak
data \cite{Abe:2008hb}. In the appendices we display the three-site model couplings
necessary to compute the one-loop flavor-dependent corrections to the $Z\to b\bar{b}$ 
decay rate, we compute the chiral current in
a general $N$-site Higgsless model, we provide a description of the
Ward-Takahashi identities that 
underlie the gaugeless limit calculation in the three-site model, and we report
the results of a unitary-gauge calculation of the process.

\section{The Three-Site Model}

\begin{figure}
\begin{center}
\includegraphics[width=8cm]{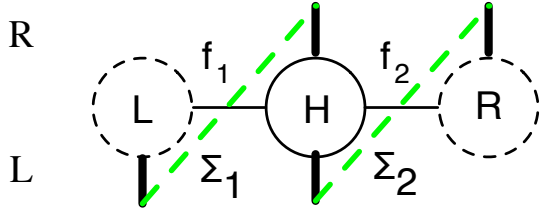}
\end{center}
\caption{The three site model in the gaugeless limit. The solid circle represents
the strong $SU(2)_H$ gauge group with coupling $\tilde{g}$, and the dashed circles 
represent global $SU(2)_{L,R}$ groups. The left-handed fermions,
denoted by the lower vertical lines and labled $\psi^{(0),(1)}_{L}$ in the text, 
are located at the first two sites, and the right-handed fermions,
denoted by the upper vertical lines and labeled $\psi^{(1)}_{R}$
and $(t^{(2)}_{R},b^{(2)}_{R})$ in the text, are located at the last two sites. The dashed lines correspond to
Yukawa couplings. We will denote the light mass-eigenstate fermion fields by $(t,b)$ and
the heavy ones by $(T,B)$. \label{fig:one}}
\end{figure}

As discussed above  we will compute
the {\it nonuniversal} correction to $Z \to b\bar{b}$ by examining the chiral current and the couplings 
of the neutral Nambu-Goldstone boson eaten by the $Z$ to 
$b$-quarks in the ``gaugeless" limit \cite{Lytel:1980zh,Barbieri:1992nz,Barbieri:1992dq,Oliver:2002up} 
of the three-site model \cite{SekharChivukula:2006cg}.
The gaugeless three-site $SU(2)_L \times SU(2)_H \times SU(2)_R$ model 
is illustrated in Fig. \ref{fig:one}, where $SU(2)_H$ is a ``hidden" gauge-symmetry
\cite{Bando:1985ej,Bando:1985rf,Bando:1988br,Casalbuoni:1985kq}
and $SU(2)_{L,R}$ are global symmetries. The
nonlinear sigma-model and gauge-theory kinetic-energy terms in 
this model are given by
\begin{equation}
{\cal L} = \sum_{i=1,2}\,\frac{f^2_i}{4}{\rm tr}\left( D^\mu \Sigma^\dagger_i D_\mu \Sigma_i\right)
-\frac{1}{4 \tilde{g}^2} (\vec{W'}^{\mu\nu})^2~,
\label{eq:threesite}
\end{equation}
where $\Sigma_1$ and $\Sigma_2$ are sigma-model fields parameterized by
\begin{equation}
\Sigma_{1,2} = \exp\left(\frac{2i\pi_{1,2}}{f_{1,2}}\right)~,
\end{equation}
where $\pi_{1,2} \equiv  \pi^a_{1,2}\sigma^a/2$, and 
where $\vec{W'}^{\mu\nu}$ is the field-strength tensor of the $SU(2)_H$ gauge-group
with gauge-fields $W'_\mu$.
The one-loop contributions described in section \ref{sec:one-loop} will
be performed in 't Hooft-Feynman gauge for the hidden $SU(2)_H$ gauge-symmetry,
and the appropriate gauge-fixing  terms (and ghost terms, though these are not needed for
the current computation) are also introduced though they are not displayed here.

The sigma-model fields transform as 
\begin{align}
\Sigma_1 & \to L\, \Sigma_1\, H^\dagger\label{eqn:xi1}~,\\
\Sigma_2 &\to H\, \Sigma_2\, R^\dagger \label{eqn:xi2}~,
\end{align}
under $SU(2)_L \times SU(2)_H \times SU(2)_R$, and hence the covariant derivatives
above are given by
\begin{align}
D^\mu \Sigma_1 = \partial^\mu \Sigma_1 +i W^{'a\mu}\Sigma_1 \frac{\sigma^a}{2}~,\\
D^\mu \Sigma_2 = \partial^\mu \Sigma_2 -  i W^{'a\mu}\frac{\sigma^a}{2} \Sigma_2~.
\end{align}
Here $f_{1,2}$ are the $f$-constants,
the analogs of $F_\pi$ in QCD, associated with the two $SU(2) \times SU(2)/SU(2)$
nonlinear sigma-models, and they
satisfy the relation
\begin{equation}
\sqrt{2} G_F = \frac{1}{v^2} =\frac{1}{f^2_1} + \frac{1}{f^2_2}\approx
\frac{1}{(250\,{\rm GeV})^2}~.
\label{eq:GF}
\end{equation}
In the gaugeless limit, the
$SU(2)_H$ vector bosons have mass
\begin{equation}
M^2_{W'} = \frac{\tilde{g}^2(f^2_1+f^2_2)}{4}~.
\end{equation}
As described in \cite{SekharChivukula:2006cg}, we get
a phenomenologically-acceptable low-energy electroweak model
if  we identify $SU(2)_L$ with the weak gauge-group and $U(1)_Y$ with the
subgroup of $SU(2)_R$ associated with $T_{3}$, and if  we work in the
limit where couplings satisfy $g_{W,Y} \ll \tilde{g}$, {\it i.e.} in
the limit $M^2_{W,Z} \ll M^2_{W'}$. 

The three-site model also incorporates the ordinary quarks and leptons, and requires the additional
heavy vectorial $SU(2)_H$ fermions that  mirror the light fermions. These
heavy Dirac fermions are the analogs of the lowest Kaluza-Klein (KK) fermion modes
which would be present in an extra-dimensional theory.
The Yukawa couplings for the third-generation are written
\begin{equation}
{\cal L}_f = -m_1 \, \bar{\psi}^{(0)}_{L} \Sigma_1 \psi^{(1)}_{R} -M  \,\bar{\psi}^{(1)}_{R}
\psi^{(1)}_{L} -  \,\bar{\psi}^{(1)}_{L} \Sigma_2
\begin{pmatrix}
m'_{t} & \\
& 0
\end{pmatrix}
\begin{pmatrix}
t^{(2)}_{R} \\
b^{(2)}_{R}
\end{pmatrix}
+ h.c.~,
\label{eq:yukawa}
\end{equation}
where we use the notation of \cite{Abe:2008hb} and we treat the
bottom-quark as massless.\footnote{Similar terms must be introduced for all
of the light quarks and leptons \protect\cite{Abe:2008hb} as well, but these terms will not play
a role in what follows.}
The transformation properties
of the fermions under $SU(2)_L \times SU(2)_H \times SU(2)_R$ are given
by
\begin{align}
\psi^{(0)}_{L} & \to L \psi^{(0)}_{L}~,\label{eq:psi0}\\
\psi^{(1)}_{R,L} & \to H \psi^{(1)}_{R,L}~, \label{eqn:psi1} \\
\begin{pmatrix}
t^{(2)}_{R} \\
b^{(2)}_{R}
\end{pmatrix}
& \to R 
\begin{pmatrix}
t^{(2)}_{R} \\
b^{(2)}_{R}
\end{pmatrix}~.
\end{align}
We will work in the limit where $M \gg m_1,\, m'_{t}$, in which the
heavy fermions are approximately the $\psi_{L1,R1}$ doublets
with a mass approximately equal to $M$. The couplings between the fermions
and the Nambu-Goldstone bosons that are necessary
for the one-loop computations are summarized in Appendix \ref{sec:threesiteappendix}.

The ratio $\epsilon_L \equiv  m_1/M$ controls the ``delocalization" 
of the left-handed
fermions, {\it i.e.} the amount to which the light left-handed mass eigenstate
fields are admixtures of fermions at the first two sites; this parameter can be
adjusted to eliminate the potentially dangerous tree-level
contributions to $\alpha S$
\cite{Cacciapaglia:2004rb,Casalbuoni:2005rs,Cacciapaglia:2005pa,Foadi:2004ps,Foadi:2005hz,Chivukula:2005bn,SekharChivukula:2005xm}. 
Therefore, at tree-level in the three-site
model, $\epsilon_L$ is taken to be flavor-universal and all of the flavor-breaking
is encoded in the values of Yukawa couplings to the right-handed fermions,
which transform under $SU(2)_R$. The three-site model at tree-level
has precisely the same flavor structure as the standard model: all of the tree-level
couplings of the left-handed fermions to the gauge bosons are flavor-diagonal
and equal, and flavor-changing neutral currents are suppressed 
\cite{SekharChivukula:2006cg}. As we will see, however, in the case of
the third generation the Yukawa couplings proportional to $m'_{t}$ will distinguish
the top- and bottom-quarks from the light generations, leading to flavor-dependence
at one loop. Before turning to this issue, however, we consider how the $W$-
and $Z$-bosons -- which, in the gaugeless limit, are treated as external fields
-- couple to the fermions and Nambu-Goldstone-bosons.

\section{Chiral Currents in the ``Gaugeless" Limit}

In the gaugeless limit, one treats the $Z$-boson
as an external field coupled to the current
\begin{equation}
j^\mu_Z = j^\mu_{3L} - j^\mu_Q \sin^2\theta_W~,
\end{equation}
with strength
\begin{equation}
g_Z =\frac{e}{\sin\theta_W \cos\theta_W}~.
\end{equation}
A crucial question\footnote{Note that the same question
does not arise for $j^\mu_Q$ which is unbroken and therefore unrenormalized. We
shall explicitly see how the difference between $j^\mu_L$ and $j^\mu_Q$ arises below.}
 is precisely what is meant by the current $j^\mu_{3L}$. We begin by computing
 this current at tree-level, which is most easily done in 
``unitary" gauge for the group $SU(2)_H$, in which the link fields 
satisfy the property that
$f_1\pi_1 = f_2 \pi_2$.
In this gauge, 
it is easy to show \cite{Hirn:2004ze}  that
\begin{equation}
 \Sigma = \Sigma_1 \cdot \Sigma_2 = \exp \left( \frac{2 i \pi}{v}\right)~,
 \end{equation}
 where
 \begin{equation}
 \frac{\pi}{v} = \frac{\pi_1}{f_1} + \frac{\pi_2}{f_2}+\ldots~,
 \end{equation}
is the non-linear sigma-model field depending only on the
massless Nambu-Goldstone bosons ($\pi_{W^\pm}$ 
and $\pi_Z$), which remain in the limit $g_{W,Y} \to 0$. 
From eqns. (\ref{eqn:xi1}) and (\ref{eqn:xi2}) we expect contributions
to $j^\mu_L$ from the $\Sigma$, which
are found to have the usual nonlinear sigma-model form \cite{Hirn:2004ze}
\begin{equation}
j^{\mu a}_{L,GB} = -i \frac{v^2}{2}\,{\rm tr} (T^a \Sigma \partial^\mu \Sigma^\dagger)~,
\label{eq:nlsm}
\end{equation}
and, from eqn. (\ref{eq:psi0}), we expect contributions of conventional form from the $\psi_{L0}$
fermions.
In addition, there are contributions
to $j^\mu_{3L}$   
proportional to the heavy $W'$ gauge-bosons  \cite{Hirn:2004ze}.
At low-momentum, the contributions proportional to the $W'$ boson
give rise to contributions to the chiral current
from the fermions $\psi_{L1,R1}$. Rather than pursue this calculation, 
we will compute the entire fermionic contribution to the 
chiral current more directly.

Consider the transformation of the sigma-model fields under a global
$SU(2)_L \times SU(2)_R$ transformation.
In unitary gauge, from eqns. (\ref{eqn:xi1}) and
(\ref{eqn:xi2}), the transformation laws of
the fields are given by 
\begin{align}
\Sigma_1 &\to L\, \Sigma_1\, H(L,R,\pi)^\dagger~, \label{eq:sigmai} \\
\Sigma_2 &\to H(L,R,\pi)\, \Sigma_2\, R^\dagger~,
\label{eq:sigmaii}
\end{align}
where $H(L,R,\pi)$ 
represents the $SU(2)_H$ gauge-transformation which, following a global $L$ and $R$
transformation, is necessary  to return to 
unitary gauge. As noted, $H(L,R,\pi)$ depends on $L$, $R$ and the pion
fields in $\Sigma_{1,2}$ as well, in such a way that
the relation $f_1 \pi_1 = f_2 \pi_2$ continues to be satisfied. 
From eqn. (\ref{eqn:psi1}) we see that  the fields $\psi_{R1,L1}$
transform through $H(L,R,\pi)$, and therefore 
transform nonlinearly under $SU(2)_L \times SU(2)_R$. In fact, the transformation
properties of $\psi_{R1,L1}$ are precisely those of matter fields in the
Callan-Coleman-Wess-Zumino (CCWZ) construction\footnote{For a recent
review connecting deconstruction and the CCWZ procedure, see \protect\cite{Thaler:2005kr}.}
 \cite{Coleman:1969sm,Callan:1969sn}.
Note, however, that the $\psi_{L0}$ and $\psi_{R2}$ fields transform linearly
under $SU(2)_L \times SU(2)_R$ -- and hence, in the computation of the current,
these fields will contribute in the conventional manner. 

Under an infinitesmal left-handed transformation
\begin{align}
\delta \psi^{(0)}_{L} & = i \alpha_L\, \psi^{(0)}_{L}~, \label{eq:f1}\\
\delta \psi^{(1)}_{R,L} & = i \eta\,  \psi^{(1)}_{R,L}~, \label{eq:f2}
\end{align}
where $\alpha_L=\alpha_L^a \sigma^a/2$ and $\eta = \eta^a \sigma^a/2$ denote
$2 \times 2$ hermitian traceless matrices, and where $\alpha^a_L$ and
$\eta^a_L$ are small parameters. From eqns. (\ref{eq:sigmai}) and
(\ref{eq:sigmaii}) we see that, to lowest order in $\alpha$, $\eta$, and $\pi_{1,2}$
\begin{align}
\frac{2\pi_1}{f_1} & \to \frac{2\pi_1}{f_1} + \alpha_L - \eta~, \\
\frac{2\pi_2}{f_2} & \to \frac{2\pi_2}{f_2} +  \eta~.
\end{align}
Imposing the relation $f_1 \pi_1 = f_2 \pi_2$, we find
\begin{equation} 
\eta = \frac{f^2_1}{f^2_1+f^2_2}\, \alpha_L + {\cal O}(\alpha^2_L, \alpha_L \pi)~,
\end{equation}
and hence, from eqns. (\ref{eq:f1}) and (\ref{eq:f2}),  the fermionic contributions to the
left-handed current are
\begin{equation}
j^{a\mu}_L = \bar{\psi}^{(0)}_{L} \frac{\sigma^a}{2} \gamma^\mu \psi^{(0)}_{L}
+ \frac{f^2_1}{f^2_1+f^2_2}\left(\bar{\psi}^{(1)}_{L} \frac{\sigma^a}{2}\gamma^\mu\psi^{(1)}_{L}
+ \bar{\psi}^{(1)}_{R} \frac{\sigma^a}{2}\gamma^\mu\psi^{(1)}_{R}\right)~.
\label{eq:leftcurrent}
\end{equation}
Note that for $f_1=f_2=\sqrt{2} v$, the fermions $\psi^{(1)}_{L,R}$ couple to the current with half the strength of $\psi^{(0)}_{L}$ -- this explains the size of weak-boson couplings of the vector (KK) fermions
in the three-site model; see eqn. (5.10) of \cite{SekharChivukula:2006cg}.\footnote{
We could have obtained
precisely the same results using the procedure of \protect\cite{Hirn:2004ze}: in 
this case, the chiral current
would contain contributions only from the pions, the fermions $\psi_{L0}$, and the $W'$-bosons.
Integrating out the $W'$ bosons, we recover the $\psi_{L1,R1}$ contributions
we find in eq. (\protect\ref{eq:leftcurrent}).} A generalization of this result to the
$N$-site global moose model appears in Appendix \ref{sec:nsite}.

The difference in the couplings of the site-0 and site-1 fermions to $j^{a\mu}_L$
is precisely the 
reason why delocalization \cite{Chivukula:2005bn}  can shift the size of the fermion couplings
to the gauge-bosons and allow -- in the case of ideal delocalization -- for $\alpha S =0$.
At tree-level, the left-handed light fermion eigenstate is given (up to corrections
of order $\epsilon^3_L$) by \cite{SekharChivukula:2006cg}
\begin{equation}
\psi_L = -\,\left(1-\frac{\epsilon^2_L}{2}\right)\psi^{(0)}_{L} + \epsilon_L \psi^{(1)}_{L}~,
\label{eq:rotation}
\end{equation}
and, hence, in the gaugeless limit the light-fermion contribution to the left-handed
current 
is given by\footnote{Eqn. (\protect\ref{eq:lightL}) is  sufficient
only to compute the leading order couplings of $j^{a\mu}_L$ to the  $Z$-boson.
Once the weak gauge-interactions are turned on,  $g_{W,Y}\neq 0$,
there are additional corrections (of order $g^2_W/\tilde{g}^2_1$) 
arising from the admixture of the site-1 gauge-boson in the light gauge-boson mass eigenstate. While
these
corrections are not included in the gaugeless limit, they are flavor universal
(to leading non-trivial order)
and do not affect the ratio $\Gamma(Z \to b\bar{b})/\Gamma(Z \to hadrons)$.}
\begin{equation}
j^{a\mu}_L \supset \left(1-\frac{\epsilon^2_L f^2_2}{f^2_1+f^2_2}\right)\bar{\psi}_L \gamma^\mu
\frac{\sigma^a}{2} \psi_L ~.
\label{eq:lightL}
\end{equation}
In addition, the rotation in eqn. (\ref{eq:rotation}) also yields couplings of the left-handed current
to mixtures of the light- and heavy-fermion eigenstates. As we will see below, these ``off-diagonal"
$Z$-boson couplings will be important in the computation of the one-loop correction to the
$Z\to b\bar{b}$ decay rate.
Note also that
there is no change in the weak-charge of the light fermions
in the limit that $f_1 \to \infty$, since in 
this case $SU(2)_L \times SU(2)_H \to SU(2)_{L+H}$ and 
the $\psi_{L1}$ couples  (see eqn. (\ref{eq:leftcurrent})) to the chiral current in the same\footnote{
This result is analogous to the GIM cancellations that occur in
the mixing of left-handed quarks  in the standard
model -- since all the quarks have the same left-handed charges, such
mixing does not result in flavor-changing neutral couplings nor does it change
the tree-level $Z$-coupling of any of the quarks.}
 way as $\psi_{L0}$.

By contrast, in the case of the unbroken electromagnetic current,
\begin{equation}
j^\mu_Q = \bar{\psi}^{(0)}_{L}Q \gamma^\mu \psi^{(0)}_{L}
+\bar{\psi}^{(1)}_{L}  Q \gamma^\mu\psi^{(1)}_{L}
+ \bar{\psi}^{(1)}_{R}  Q \gamma^\mu\psi^{(1)}_{R}
+ \bar{\psi}^{(2)}_{R}  Q \gamma^\mu\psi^{(2)}_{R}~,
\end{equation}
where $Q=diag(2/3, -1/3)$ is the quark charge-matrix. In this
case all fermions couple to the photon
in the same way, and fermion delocalization (mixing) cannot
change the electric charge of the fermions.

\section{One-Loop Corrections}
\label{sec:one-loop}

\begin{figure}
\begin{center}
\includegraphics[width=4cm]{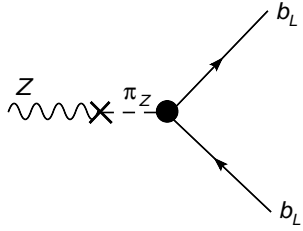}
\end{center}
\caption{The Nambu-Goldstone boson contributions to the current $j^\mu_{3L}$
give rise to corrections to the $Zb\bar{b}$ vertex  \protect\cite{Barbieri:1992nz,Barbieri:1992dq}. 
In this figure, the black circle represents vertices in the effective action of
the form $\partial^\mu \pi_Z \bar{b}_L \gamma_\mu b_L$ .\label{fig:two}}
\end{figure}

As previously noted, in the gaugeless limit, one computes
the couplings of external $Z$ bosons to the current $j^\mu_{3L} - j^\mu_Q 
\sin^2 \theta_W$. The current $j^\mu_Q$ is conserved and
therefore unrenormalized and flavor-universal. 
The flavor non-universal couplings to the $Z$ boson occur
because $j^\mu_{3L}$ is a current that corresponds to a spontaneously broken symmetry,
and therefore arise from the non-linear sigma model currents in eqn. (\ref{eq:nlsm}).
As we will show below, there are flavor-dependent contributions in the effective
action to operators of the form
\begin{equation}
A\, \partial^\mu\pi_Z \bar{b}_L \gamma_\mu b_L~.
\label{eq:operator}
\end{equation}
Through the diagram  illustrated in Fig. \ref{fig:two}, such an
operator shifts\footnote{
This is, essentially, a diagrammatic interpretation of the Ward-Takahashi Identity argument
given in  \protect\cite{Barbieri:1992nz,Barbieri:1992dq}; see Appendix \protect\ref{sec:wi}.} the
left-handed $Zb\bar{b}$ coupling to
\begin{equation}
g_Z \( -\frac{1}{2} + \delta g^{b\bar{b}}_L + \frac{1}{3} \sin^2\theta_W\)~,
\label{eq:defdeltag}
\end{equation}
where $g_Z$ and $\sin^2\theta_W$ are flavor-independent, and
where the flavor-dependent correction is
\begin{equation}
\delta g^{b\bar{b}}_L = \frac{v}{2}\,A~.
\end{equation}

\begin{figure}
\begin{center}
\includegraphics[width=4cm]{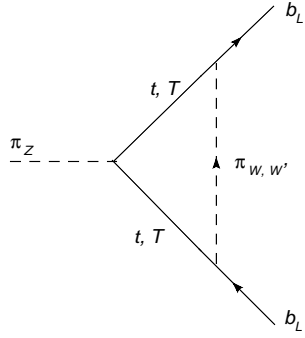}
\end{center}
\caption{The flavor-dependent vertex corrections to the $Z \to b\bar{b}$ branching
ratio, as computed in the gaugeless limit of the three-site model.
Here $\pi_Z$ and $\pi_W$ denote the neutral and charged 
Nambu-Goldstone bosons which remain in the gaugeless limit. We perform this
calculation in 't Hooft-Feynman gauge for $SU(2)_H$, and the $\pi_{W'}$ denote the
unphysical Nambu-Goldstone bosons eaten by the heavy $W'$ bosons; the flavor-dependent
contributions from the $W'$ gauge-bosons are suppressed by an additional
factor of $M^2_{W'}/M^2$. In this diagram we denote the heavy Dirac partners
of the top-quark by $T$.
The contribution with intermediate $t$ quarks and exchange of the $\pi_W$ boson
yields the usual standard
model one-loop correction \protect\cite{Barbieri:1992nz,Barbieri:1992dq}. 
The other contributions are new in the three-site model.\label{fig:three}}
\end{figure}

\begin{figure}
\begin{center}
\includegraphics[width=3cm]{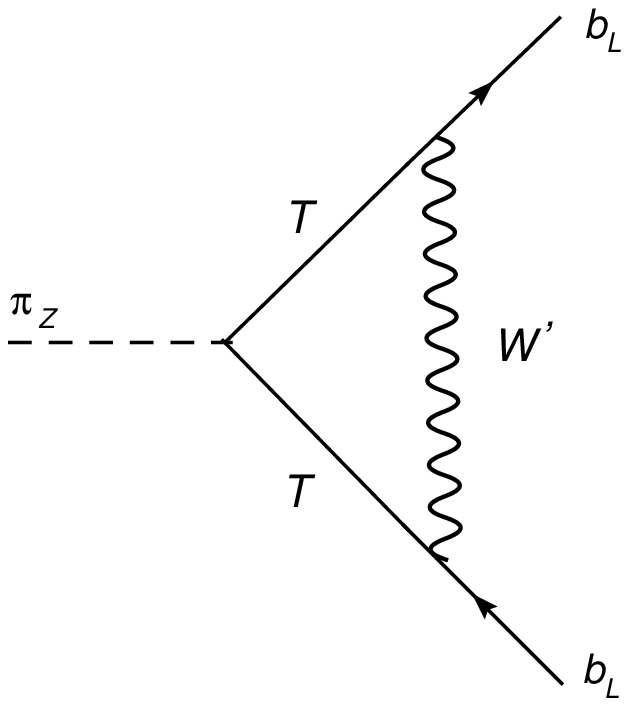}
\includegraphics[width=6.75cm]{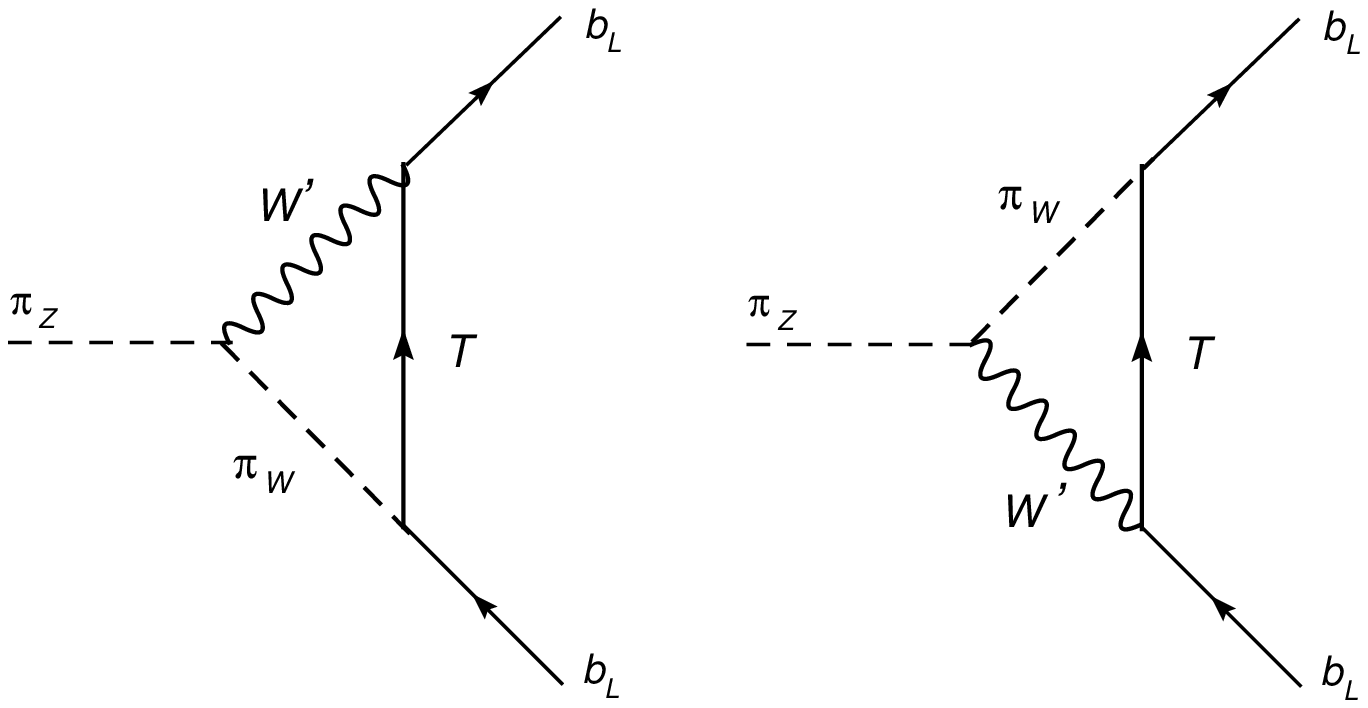}
\includegraphics[width=7.25cm]{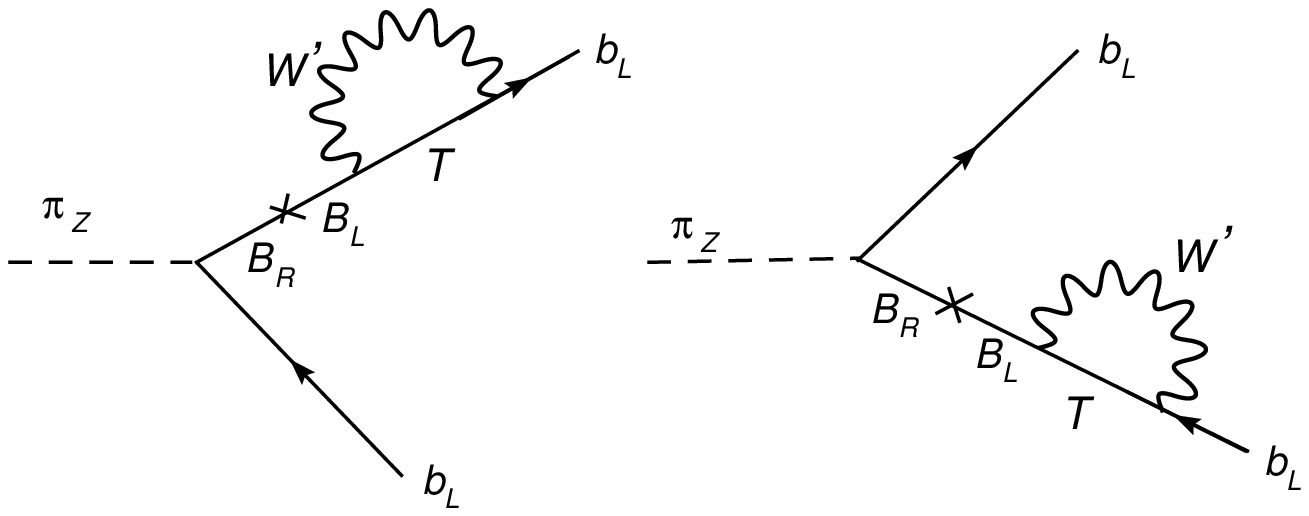}
\end{center}
\caption{These $W'$-exchange diagrams yield  flavor non-universal
contributions which are suppressed, relative to the contributions we have kept, by  $M^2_{W'}/M^2$.
\label{fig:five}
}
\end{figure}

The vertex diagrams (in 't Hooft-Feynman gauge) leading to flavor-dependent 
contributions\footnote{We neglect here, for example, vertex diagrams
involving the exchange of $\pi_{Z,Z'}$ and intermediate $b$ or $B$ quarks: in
the limit in which we ignore the $b$-quark mass, these contributions
are the same for all flavors of quarks. Also, in 't Hooft-Feynman gauge the flavor-dependent contributions
arising from $W'$ exchange are suppressed since the heavy Dirac partners
of all of the fermions are nearly degenerate. As we discuss at the end of this section,
there are subleading -- suppressed by $M^2_{W'}/M^2$ -- contributions from
the additional diagrams  in figure \protect\ref{fig:five}.}
to the operator in eqn. (\ref{eq:operator})
are illustrated in Fig. \ref{fig:three}.
The triangle contribution with $\pi_W$-exchange and two intermediate top-quarks yields the usual
gaugeless standard model correction\footnote{There are additional one-loop corrections 
\protect\cite{Akhundov:1985fc,Beenakker:1988pv,Bernabeu:1987me} proportional
to weak couplings (and at most logarithmically dependent on $m_t$) which cannot be computed
in the gaugeless limit.}  \cite{Barbieri:1992nz,Barbieri:1992dq,Oliver:2002up}
\begin{equation}
(\delta g^{b\bar{b}}_L)_{sm} = \frac{m^2_t}{16 \pi^2 v^2}~.
\label{eq:zbbsm}
\end{equation}
On the other hand, the triangle diagrams including contributions from the heavy Dirac
partners of the top-quarks and/or the exchange of the $\pi_{W'}$ boson,
yield the correction
\begin{equation}
(\delta g^{b\bar{b}}_L)_{3-site\,vertex} =
-\,\frac{1}{2(4\pi)^2} \frac{m^2_t}{v^2} \frac{f^2_1 f^2_2}{(f^2_1+f^2_2)^2} 
\log\left(\frac{M^2}{m^2_t}\right)~,
\label{eq:zbb3site}
\end{equation}
where we have used eqn. (\ref{eq:GF}) and the relation
\begin{align}
m_t & \approx \frac{m_1 m'_t}{M}~.\label{eq:mt}
\end{align}

\begin{figure}
\begin{center}
\includegraphics[width=9cm]{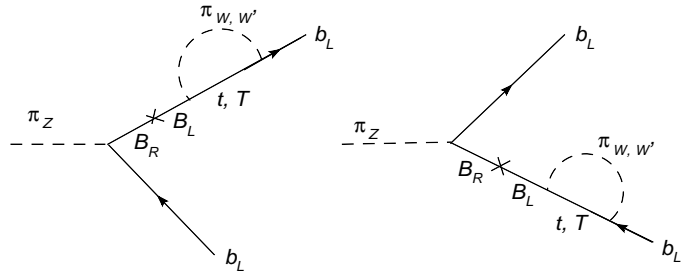}
\end{center}
\caption{Wavefunction mixing contributions to $Z\to b\bar{b}$ in
the three site model. The $B$ fermions are the heavy Dirac
partners of the bottom-quark.  \label{fig:four}}
\end{figure}

In addition to the vertex corrections, there are flavor-dependent wavefunction mixing 
contributions\footnote{Again, as in the case of the vertex diagrams, 
the flavor-dependent contributions
from the diagrams with $W'$ exchange in figure \ref{fig:five} are suppressed.}  which
must be added, as illustrated in Fig. \ref{fig:four}. These contributions exist in the three-site
model because of the existence of  the $\pi_Z \bar{b}_L B_R$ vertex, which couples
the $b$ to its heavy Dirac partner $B$ (see eqn. (\ref{eq:bBvertex})).
There is no analogous contribution in the standard model, because the $\pi_Z \bar{b}_L b_R$
vertex vanishes in the limit of zero bottom-quark mass. In Appendix \ref{sec:wi}, we show how
these terms arise from the Ward-Takahashi identity of the three-site model. Here we present
an alternative derivation to show directly how
these extra contributions give rise to operators of the form in eqn. (\ref{eq:operator}) after
integrating out the heavy $B$ field. Including the wavefunction mixing, we may write the
effective action as
\begin{equation}
{\cal L} = \bar{b}_L i\slashi{\partial} b_L + \bar{B} i\slashi{\partial} B - M \bar{B} B
+ \eta \bar{B}_L i\slashi{\partial}b_L + ig_{\pi b B} \pi_Z \bar{b}_L B_R +{h.c.}+\ldots~,
\label{eq:action}
\end{equation}
where $\eta$ is a small-parameter of one-loop order. The linear terms in the equations
of motion for the $B$ field are then
\begin{align}
i \slashi{\partial} B_L - M B_R + i\eta\slashi{\partial}b_L+\ldots & =0~,\\
i \slashi{\partial} B_R - M B_L +\ldots &=0~,
\end{align}
where the ellipses refer to (interaction) terms with more than one field. Integrating out
the $B$ field in the large-$M$ limit, we find\footnote{An alternative procedure would
be to diagonalize the kinetic energy and mass terms in (\protect\ref{eq:action}).
In this case the analysis is a bit more complicated, though the $S$-matrix
that arises is equivalent. Note that on-shell matrix elements of the operator
in eqn. (\protect\ref{eq:operator}) are of order $m_b$. One must include a non-zero mass
for the bottom-quark and carefully keep track of terms of order $m_b$; doing so, one finds
that $B_R$ mixes with $b_R$, and yields an effect equivalent to the one we compute above. Note
that although one must keep terms proportional to $m_b$ in this procedure, the final
correction to the $Z\to b\bar{b}$ amplitude is not proportional to $m_b$.}
\begin{equation}
B_R = i\,\frac{\eta}{M}\slashi{\partial} b_L +\ldots~.
\end{equation}
Plugging this expression into the $\pi_Z \bar{b}_L B_R$ (and Hermitian conjugate) 
coupling yields the operator in
eqn. (\ref{eq:operator}). The wavefunction mixing diagrams are
logarithmically divergent. Performing the calculation using dimensional regularization
and  using $\overline{\rm MS}$ we find that
 the wavefunction mixing contributions then yield
\begin{equation}
(\delta g^{b\bar{b}}_L)_{3-site\,wavefunction} =
+\,\frac{1}{2(4\pi)^2} \frac{m^2_t}{v^2} \frac{f^2_1 f^2_2}{(f^2_1+f^2_2)^2} 
\left(\log\left(\frac{\mu^2}{m^2_t}\right)+\frac{3}{2}\right)~,
\label{eq:zbb3sitewf}
\end{equation}
where $\mu$ is the regularization mass.

Adding all of the flavor-dependent contributions from eqns. (\ref{eq:zbbsm}), (\ref{eq:zbb3site})
and (\ref{eq:zbb3sitewf}), we find the total contribution
\begin{equation}
\delta g^{b\bar{b}}_L= \frac{m^2_t}{(4\pi)^2 v^2} \left[1+
\frac{f^2_1 f^2_2}{2(f^2_1+f^2_2)^2} 
\left(\log\left(\frac{\mu^2}{M^2}\right)+\frac{3}{2}\right)+\delta g^{b\bar{b}}_L(\mu)\right]~,
\label{eq:result}
\end{equation}
where $\delta g^{b\bar{b}}_L(\mu)$ represents the three-site model counterterm,
renormalized at scale $\mu$,  required
to renormalize the theory appropriately.  As shown in \cite{Abe:2008hb},
and discussed further in the next section,
at one-loop there are flavor-dependent renormalizations of the heavy Dirac
masses $M$ in eqn. (\ref{eq:yukawa}), and $\delta g^{b\bar{b}}_L(\mu)$ represents
the effect of the counterterm necessary to implement this renormalization.

Two properties of this result are worth commenting on. First, note that the
additional three-site contributions vanish in the limit that $f_1\ {\rm or }\ f_2 \to \infty$ with
$v$ held fixed.
This is reasonable since the three-site model reduces to the electroweak chiral
Lagrangian 
\cite{Appelquist:1980ae,Appelquist:1980vg,Longhitano:1980iz,Longhitano:1980tm,Appelquist:1993ka}
in this limit, and the $Z\to b\bar{b}$ corrections must therefore reduce
to those of the standard model. Second, note that the corrections proportional
to $\log(m_t)$ cancel when we add the three-site vertex and wavefunction
mixing contributions -- this confirms the effective field-theory
argument given in  \cite{SekharChivukula:2006cg}, which noted that once the
heavy fermions were integrated out there were no operators in the effective theory 
whose scaling could affect the size of the $Zb\bar{b}$ coupling.

The results described above are the 
leading-order flavor-dependent contributions, in 't Hooft-Feynman gauge for
the hidden $SU(2)_H$ gauge theory, 
arising from the diagrams in figures \ref{fig:three} and \ref{fig:four}; there
are also subleading diagrams as illustrated in fig. \ref{fig:five}. We
have checked our results by computing the one-loop
corrections to $Z \to b\bar{b}$ in the full three-site model 
in unitary gauge, as described in Appendix \ref{sec:unitary}. 

\section{RGE Analysis}

It is interesting to see how the logarithmic term in eqn. (\ref{eq:result}) 
can be reproduced using the RGE analysis of ref. \cite{Abe:2008hb}. In this
case, one analyzes the three-site model in the limit $\Lambda \gg M \gg M_{W'}$
in effective field theory. 
As described in  \cite{Abe:2008hb}, we define the parameters in eqns. (\ref{eq:threesite})
and (\ref{eq:yukawa}) in terms of their values at the cutoff scale $\Lambda$.
For the energy regime between $\Lambda$ and $M$, one
considers the one-loop running of the operators in the full three-site model. At
scale $M$, one ``integrates out" the heavy fermions and constructs an effective theory
with  one-site delocalization \cite{Chivukula:2005bn}. Subsequently, one computes
the running of the operators from the scale $M$ to the scale $M_{W'}$. At this
stage, one integrates out the heavy $W'$ fields and matches the theory to
the electroweak chiral lagrangian 
\cite{Appelquist:1980ae,Appelquist:1980vg,Longhitano:1980iz,Longhitano:1980tm,Appelquist:1993ka}.
Lastly, one runs in the electroweak chiral lagrangian to energies of order $M_Z$
to analyze the electroweak processes of interest.

The only flavor-dependent terms arise from the Yukawa couplings proportional
to $m'_t$. As shown in ref. \cite{Abe:2008hb}, these terms give rise
at one-loop to a flavor-dependent renormalization of the masses of the heavy
Dirac fermions in
the running between the cutoff scale $\Lambda$ and the heavy fermion scale
$M$. This flavor-dependent
renormalization of the $\psi^{(1)}_L$ wavefunction arises from the $\pi_2$
interactions  eqn. (\ref{eq:yukawa}). Conventionally normalizing the third-generation
$\psi^{(1)}_L$ fields then leads to a flavor-dependent shift in the Dirac mass $M$,
and the RGE running of this parameter.

The quantity of phenomenological interest is the ratio $\epsilon_L = m_1/M$,
which defines the delocalization of the light left-handed fermions.  We find \cite{Abe:2008hb}
the running
\begin{equation}
  \mu \dfrac{d}{d\mu} \left(
    \dfrac{m_1}{M}
  \right)_{3rd}
  = \dfrac{1}{(4\pi)^2} \dfrac{m_1}{M}\left(
    \dfrac{9}{2} \tilde{g}^2 - \dfrac{9}{2}
    \dfrac{m_1^2}{f_1^2}-\dfrac{m_t'^2}{f_2^2} 
    \right),
\label{eq:rge3}
\end{equation}
for the third generation of quarks, while
\begin{equation}
  \mu \dfrac{d}{d\mu} \left(
    \dfrac{m_1}{M}
  \right)_{1st}
  = \dfrac{1}{(4\pi)^2} \dfrac{m_1}{M}\left(
    \dfrac{9}{2} \tilde{g}^2 - \dfrac{9}{2}
    \dfrac{m_1^2}{f_1^2}
    \right)
\label{eq:rge1}
\end{equation}
for the first and second generations of quarks.
Assuming universality at the cutoff scale $\Lambda$
\begin{equation}
  \left. \left(
    \dfrac{m_1}{M}
  \right)_{1st} \right|_{\mu=\Lambda}
 =\left. \left(
    \dfrac{m_1}{M}
  \right)_{3rd} \right|_{\mu=\Lambda},
  \label{eq:universal}
\end{equation}
we find that the renormalization group equations
Eqs.(\ref{eq:rge3})-(\ref{eq:rge1}) induce a flavor-non-universal
correction 
\begin{equation}
  \Delta \epsilon^2_L
  \equiv
  \left. \left(
    \dfrac{m_1}{M}
  \right)_{3rd}^2 \right|_{\mu=M}
 -\left. \left(
    \dfrac{m_1}{M}
  \right)_{1st}^2 \right|_{\mu=M}
  =
   \dfrac{1}{(4\pi)^2}\dfrac{m_t'^2}{f_2^2}\left(\dfrac{m_1}{M}\right)^2
   \ln\dfrac{\Lambda^2}{M^2},
\label{eq:key}
\end{equation}
at the KK fermion mass scale $M$. 
Below the scale $M$ all subsequent evolution is flavor-universal.

The effect of this flavor-dependent renormalization
is to shift the value of $\epsilon_L$ for the third-generation quarks
\begin{equation}
\epsilon^2_L \to \epsilon^2_L+ \frac{1}{(4\pi)^2} \frac{m^2_t}{f^2_2} 
\log\frac{\Lambda^2}{M^2}~,
\end{equation}
where we have simplified by using eqn. (\ref{eq:mt}).
Using eqns.  (\ref{eq:GF}),  (\ref{eq:lightL}), and (\ref{eq:defdeltag}), we then find the correction
\begin{equation}
(\delta g^{b\bar{b}}_L)_{3-site} =
+\,\frac{1}{2(4\pi)^2} \frac{m^2_t}{v^2} \frac{f^2_1 f^2_2}{(f^2_1+f^2_2)^2} 
\log\left(\frac{\Lambda^2}{M^2}\right)~.
\label{eq:zbb3sitewfrge}
\end{equation}

The result of eqn. (\ref{eq:zbb3sitewfrge}) is in agreement with the second term in eqn. (\ref{eq:result}) with the identifications
\begin{align}
\log \Lambda^2 & = \log{\mu}^2 +\frac{3}{2}~, \\
\delta g^{b\bar{b}}_L(\Lambda) & = 0~.
\end{align}
From the discussion above, we see that the $\delta g^{b\bar{b}}_L(\mu)$ represents
the dependence of the $Z\to b\bar{b}$ amplitude on the (flavor-dependent) heavy
Dirac mass, and the condition $\delta g^{b\bar{b}}_L(\Lambda)=0$ expresses the 
choice (eqn. (\ref{eq:universal})) of flavor-universal Dirac masses when the theory is renormalized at the
cutoff scale $\Lambda$.

\section{Limits from $R_b$}

The best way to compare our expressions
(eqn. (\ref{eq:zbb3sitewfrge})) 
for the shifted $Zb\bar{b}$ coupling to experiment is by calculating the ratio $R_b$:
\begin{equation}
  R_b \equiv \dfrac{\Gamma(Z\to b\bar{b})}{\Gamma(Z\to \mbox{hadrons})}.
\end{equation}
This ratio can be evaluated as 
\begin{equation}
  R_b \simeq \dfrac{g_{bL}^2+g_{bR}^2}
             {\sum_{f=u,d,c,s,b} (g_{fL}^2+g_{fR}^2)} ,
\end{equation}
where $g_{fL}$ ($g_{fR}$) denotes the $Z$ boson coupling to the left
(right) handed $f$ quark. Following Ref. \cite{Oliver:2002up}, we may decompose the $Z b_L \bar{b}_L$
couplings (as elsewhere in this paper) into standard model and New Physics pieces,
\begin{equation}
  g_{bL} = g_{bL}^{\rm SM} + \delta g_L^{\rm NP}.
\end{equation}
We may express the New Physics effect on $R_b$ in terms
of $\delta g_L^{\rm NP}$ \cite{Oliver:2002up}
\begin{equation}
  \delta R_b = 2 R_b (1-R_b) \dfrac{g_{bL}}{g_{bL}^2+g_{bR}^2}
  \delta g_L^{\rm NP}.
\label{eq:rbnew}
\end{equation}
To leading order we may  simplify this expression by inserting the value of 
 $R_b$ predicted by the standard model \cite{Amsler:2008zz}
\begin{equation}
  R_b^{\rm SM} = 0.21584 \pm 0.00006.
\label{eq:rbsm}
\end{equation}
as well as the standard model values\footnote{The level of
accuracy for $\sin^2\theta_W$ in eqn. (\protect\ref{eq:gbtree}) is sufficient
to compute the size of the one-loop correction to $R_b$ to the required accuracy.}
 of $g_{bL}$ and $g_{bR}$ 
\begin{equation}
  g_{bL} = -\dfrac{1}{2} + \dfrac{1}{3}\sin^2\theta_W, \qquad
  g_{bR} = \dfrac{1}{3}\sin^2\theta_W, \qquad  \sin^2\theta_W \simeq 0.23, 
\label{eq:gbtree}
\end{equation}
to obtain
\begin{equation}
  \delta R_b \simeq -0.774 \times \delta g_L^{\rm NP}.
\label{eq:rbnew2}
\end{equation}

In order to place a limit on the size of the coupling shift due to New Physics, we compare the result above with the data.  The observed value of  $R_b$ \cite{Amsler:2008zz} is
\begin{equation}
  R_b^{\rm obs} = 0.21629 \pm 0.00066~,
\label{eq:rbobs}
\end{equation}
and the gap between this and the standard model value (\ref{eq:rbobs}) is
\begin{equation}
  \delta R_b^{\rm obs} \equiv R_b^{\rm obs}-R_b^{\rm SM}
        = (4.5 \pm 6.6)\times 10^{-4}.
\label{eq:rbobs2}
\end{equation}
Comparing this with eqn. (\ref{eq:rbnew2}) yields the general constraint
\begin{equation}
  \delta g_L^{\rm NP} = (-5.8\pm   8.6)\times 10^{-4}~,
\label{eq:bound1}
\end{equation}
on the shift any New Physics may induce in the $Zb\bar{b}$ coupling.

For the three-site model in particular, $\delta g_L^{\rm NP}$ corresponds to the coupling shift we obtained in eqn. (\ref{eq:zbb3sitewfrge}):
\begin{equation}
  \delta g_L^{\rm NP}
  = \dfrac{m_t^2}{(4\pi)^2 v^2} F 
  = \dfrac{\sqrt{2} G_F m_t^2}{(4\pi)^2} F, 
\end{equation}
with
\begin{equation}
  F = \dfrac{f_1^2 f_2^2}{2(f_1^2+f_2^2)^2} \ln \dfrac{\Lambda^2}{M^2}.
\end{equation}
If we insert the values 
\begin{equation}
  m_t \simeq 171\mbox{GeV}, \qquad
  G_F \simeq 1.166\times 10^{-5}\mbox{GeV}^{-2}~,
\end{equation}
we obtain the following  bound on $F$ 
\begin{equation}
  F = -0.19 \pm 0.28.
\label{eq:bound2}
\end{equation}
Since $F$ is theoretically constrained to be positive,
we need to be a bit careful when we deduce a 95\% CL bound on
$F$.
Following \cite{Oliver:2002up}, we apply the method proposed in
\cite{Feldman:1997qc}, including the information in its Table X; our result is
\begin{equation}
  F < 0.38 \qquad (\mbox{95\%\ CL}).
\end{equation}
From this, we may derive a bound on $\Lambda/M$.
For example, in the case   $f_1 = f_2=\sqrt{2} v$
(which corresponds to maximal unitarity delay \cite{SekharChivukula:2006cg}), we find
\begin{equation}
  \dfrac{\Lambda}{M} < 4.6 
 \qquad (\mbox{95\%\ CL}).
\end{equation}
We expect a $\Lambda$ of order 4 TeV or less, since $\Lambda \leq 4\pi f_{1,2} \approx 4\sqrt{2}\pi v$, 
and therefore we obtain the constraint that the heavy fermion masses should 
be at least of order one TeV -- a constraint satisfied
automatically by the three-site model in the range of parameter space allowed by other precision electroweak
data \cite{SekharChivukula:2006cg,Abe:2008hb,Sekhar Chivukula:2007mw}.

\section{Conclusions}

We have computed the  flavor-dependent chiral-logarithmic corrections to the decay
$Z \to b\bar{b}$ in the three site Higgsless model and have demonstrated that the diagramatic calculation in the gaugeless limit agrees with an RGE analysis of the effective theory.  
We have shown the necessity of carefully incorporating the effects of mixing between
the light- and heavy-fermions in the computation of this result; such effects are
not present in many other 
theories beyond the Standard Model,  
such as the MSSM or models featuring extended electroweak gauge groups  
but no new fermions. Comparing our three-site model result
\begin{equation}
\delta g^{b\bar{b}}_L= \frac{m^2_t}{(4\pi)^2 v^2} \left[1+
\frac{f^2_1 f^2_2}{2(f^2_1+f^2_2)^2} 
\log\left(\frac{\Lambda^2}{M^2}\right)\right]~,
\nonumber
\end{equation}
with the data on $R_b$ yields the rather mild constraint that the heavy fermions have masses of at least 1 TeV.  This limit is automatically satisfied by the three-site model in the range of parameter space allowed by other precision electroweak data \cite{SekharChivukula:2006cg,Abe:2008hb,Sekhar Chivukula:2007mw}. Moreover, the form we obtain for the chiral currents in an $N$-site global
moose model with fermion delocalization suggests that the effects on $R_b$ in such models
(and therefore in continuum models as well) will be similar.

It is interesting to note the contrast between our results and those for Higgsless models without delocalization (those in which the $W'$ is not fermiophobic).  We found that in the three-site model, corrections proportional to $\ln{m_t}$ cancel between the vertex and wavefunction mixing contributions.   Since the effective theory is valid only to $\Lambda \approx 4\pi\sqrt{2}v \approx {\rm 4\, TeV}$, while precision electroweak data force $M$ to lie above 1.8 TeV or so \cite{SekharChivukula:2006cg,Abe:2008hb,Sekhar Chivukula:2007mw}, the remaining chiral log factor $\ln(\Lambda^2 / M^2)$ cannot be large.  However, in generic extra dimensional models when one integrates out the  KK modes \cite{Oliver:2002up}, corrections proportional to $\ln( M_{W'}^2 / m_t^2)$ can persist and lead to more stringent experimental constraints.

\acknowledgements

The figures in this paper were generated using the program
Jaxodraw \cite{Binosi:2003yf}, and many of the one-loop calculations
were completed with the aid of FeynRules \cite{Christensen:2008py}.
T.A. is supported in part by the JSPS Grant-in-Aid No.204354 and 
  Nagoya University Global COE program, Quest for Fundamental
  Principles in the Universe.  He thanks the Michigan State
  University high-energy theory group for their hospitality during
  the completion of this work.
 M.T.Õs work is supported in
part by the JSPS Grant-in-Aid for Scientific Research
No. 20540263.
R.S.C., E.H.S., N.C., and K.H. are supported in part by the US National Science Foundation under
grant  PHY-0354226. S.M. is supported by the U.S. Department of Energy under
Grant No. DE-FG02-06ER41418.

\appendix

\section{Three-Site Model Couplings, Masses, and Eigenstates}

\label{sec:threesiteappendix}

In this appendix we present the couplings, masses, and mass eigenstate
fields required in the computation of $Z\to b\bar{b}$ in the gaugeless limit.
The quark mass matrices of this model are
\begin{equation}
  (\bar{b}_L^{(0)}, \bar{b}_L^{(1)})
  \left(
    \begin{array}{cc}
      m_1 & 0 \\
      M   & 0 
    \end{array}
  \right) \left(
    \begin{array}{c}
      b_R^{(1)} \\
      b_R^{(2)}
    \end{array}
  \right),
  \qquad
  (\bar{t}_L^{(0)}, \bar{t}_L^{(1)})
  \left(
    \begin{array}{cc}
      m_1 & 0 \\
      M   & m_t' 
    \end{array}
  \right) \left(
    \begin{array}{c}
      t_R^{(1)} \\
      t_R^{(2)}
    \end{array}
  \right)~.
\end{equation}
Denoting the heavy and light mass eigenstates by $(t,b)$ and $(T,B)$, we
diagonalize the matrices by
\begin{equation}
  \left(
    \begin{array}{c}
      b_L \\
      B_L
    \end{array}
  \right)
  = V_{bL}\left(
    \begin{array}{c}
      b_L^{(0)} \\
      b_L^{(1)}
    \end{array}
    \right), \qquad
    V_{bL} \simeq \left(
      \begin{array}{cc}
       - 1 & \dfrac{m_1}{M} \\
         - \dfrac{m_1}{M} & - 1
      \end{array}
    \right),
\end{equation}
\begin{equation}
  \left(
    \begin{array}{c}
      b_R \\
      B_R
    \end{array}
  \right)
  = V_{bR}\left(
    \begin{array}{c}
      b_R^{(1)} \\
      b_R^{(2)}
    \end{array}
    \right), \qquad
    V_{bR} = \left(
      \begin{array}{cc}
        0 & 1 \\
        -1 & 0 
      \end{array}
    \right),
\end{equation}
\begin{equation}
  \left(
    \begin{array}{c}
      t_L \\
      T_L
    \end{array}
  \right)
  = V_{tL}\left(
    \begin{array}{c}
      t_L^{(0)} \\
      t_L^{(1)}
    \end{array}
    \right), \qquad
    V_{tL} \simeq \left(
      \begin{array}{cc}
        - 1 & \dfrac{m_1}{M} \\
        - \dfrac{m_1}{M} & - 1
      \end{array}
    \right),
\end{equation}
and
\begin{equation}
  \left(
    \begin{array}{c}
      t_R \\
      T_R
    \end{array}
  \right)
  = V_{tR}\left(
    \begin{array}{c}
      t_R^{(1)} \\
      t_R^{(2)}
    \end{array}
    \right), \qquad
    V_{tR} \simeq \left(
      \begin{array}{cc}
        -\dfrac{m_t'}{M} & 1 \\
        -1 & -\dfrac{m_t'}{M} 
      \end{array}
    \right).
\end{equation}
Note that we approximate the $b$ quark as massless, which will be
sufficient for the computation of the one-loop corrections to the vertex and
wavefunction-mixing diagrams.
Assuming $M \gg m_1, m_t'$, the $t$ quark mass is evaluated as
\begin{equation}
  m_t \simeq \dfrac{m_1 m_t'}{M}.
  \label{eq:mtop}
\end{equation}
The heavy KK quarks $T$ and $B$ are almost degenerate
\begin{equation}
  m_T \simeq m_B \simeq M.
\end{equation}

We next turn to the Nambu-Goldstone boson sector.
We define $\pi_W$ and $\pi_{W'}$ as
\begin{equation}
  \left(
    \begin{array}{c}
      \pi_{W} \\
      \pi_{W'}
    \end{array}
  \right)
  = V_{\pi}\left(
    \begin{array}{c}
      \pi_1 \\
     \pi_2
    \end{array}
    \right), \qquad
    V_{\pi} = \dfrac{1}{\sqrt{f_1^2+f_2^2}}
      \left(
      \begin{array}{cc}
        f_2 & f_1 \\
       -f_1 & f_2
      \end{array}
    \right).
\end{equation}
The $\pi_{W'}$ bosons are eaten by the gauge boson $W'^\mu$, while the $\pi_W$
bosons are physical and remain massless.  As appropriate, we will denote the
neutral Nambu-Goldstone boson, which will be eaten by the $Z$ when $g_{W,Y} \neq 0$,
 by $\pi_Z$.

We are now ready to evaluate various coupling strengths that are
relevant for the one-loop computation of the $Zb\bar{b}$ vertex in the
gaugeless limit,
\begin{eqnarray}
  g_{\bar{b}_L t_R \pi_W}
  &\simeq& \sqrt{2}\dfrac{\sqrt{f_1^2+f_2^2}}{f_1f_2} 
            \dfrac{m_1 m_t'}{M}
  \simeq \sqrt{2} \dfrac{m_t}{v},
\end{eqnarray}
\begin{eqnarray}
  g_{\bar{b}_L T_R \pi_W}
  &\simeq& \sqrt{2} \dfrac{m_1 f_2}{f_1 \sqrt{f_1^2+f_2^2}}
 ~, \\
 g_{\bar{b}_L t_R \pi_{W'}}
   &\simeq& 0,\\
  g_{\bar{b}_L T_R \pi_{W'}}
  &\simeq& -\dfrac{\sqrt{2} m_1}{\sqrt{f_1^2+f_2^2}}
           ~,\\
   g_{\bar{b}_L B_R \pi_Z}
  &\simeq& -\dfrac{m_1 f_2}{f_1\sqrt{f_1^2+f_2^2}}\label{eq:bBvertex} ~,\\
g_{\bar{B}_L t_R \pi_W}
    &\simeq& \sqrt{2} \dfrac{m_t' f_1}{f_2 \sqrt{f_1^2+f_2^2}}
  ~, \\
  g_{\bar{B}_L T_R \pi_W}
    &\simeq& 
  \dfrac{\sqrt{2}\,M}{\sqrt{f_1^2+f_2^2}}
  \left[
    \dfrac{f_2}{f_1}\dfrac{m_1^2}{M^2} 
  + \dfrac{f_1}{f_2} \dfrac{m_t'^2}{M^2}
  \right],\label{eq:bigi}\\
  g_{\bar{B}_L t_R \pi_{W'}}
  &\simeq& -\dfrac{\sqrt{2} m_t'}{\sqrt{f_1^2+f_2^2}}
            ~,\\
  g_{\bar{B}_L T_R \pi_{W'}}
  &\simeq& \dfrac{\sqrt{2}\,M}{\sqrt{f_1^2+f_2^2}}
            \left[
              -\dfrac{m_1^2}{M^2}
              +\dfrac{m_t'^2}{M^2}
            \right],\label{eq:bigii}\\
  g_{\bar{t}_L t_R \pi_Z}
  &\simeq& \dfrac{\sqrt{f_1^2+f_2^2}}{f_1 f_2} 
            \dfrac{m_1 m_t'}{M}
  \simeq \dfrac{m_t}{v},\\
  g_{\bar{T}_L t_R \pi_Z}
  &\simeq& -\dfrac{m_t' f_1}{f_2 \sqrt{f_1^2 + f_2^2}}
 ~,\\
    g_{\bar{T}_L T_R \pi_Z} &\simeq&
    \dfrac{M}{\sqrt{f^2_1+f^2_2}}
  \left[ \dfrac{f_1\,{m^\prime_t}^2}{f_2 M^2}
  + \frac{f_2\, m^2_1}{f_1\, M^2}\right]~. \label{eq:bigiii}
\end{eqnarray}
In these expressions we have ignored terms of ${\cal O}(m_1/M)^2$
and ${\cal O}(m_t'/M)^2$; note that the couplings in eqns. (\ref{eq:bigi}) ,
(\ref{eq:bigii}), and (\ref{eq:bigiii})  are enhanced by the (potentially large)
factor $(M/\sqrt{f^2_1+f^2_2})$.

\section{Chiral Currents in the $N$-site Model}
\label{sec:nsite}

\begin{figure}
\begin{center}
\includegraphics[width=12cm]{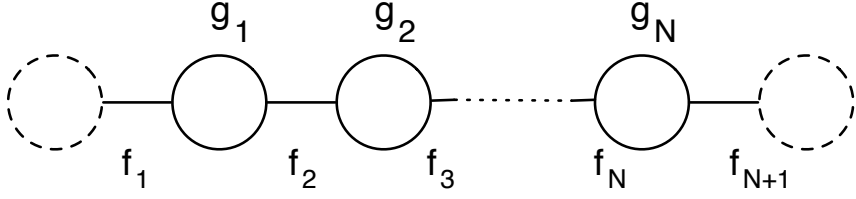}
\end{center}
\caption{The $N$-site global moose. The global-symmetries are at
sites $0$ and $N+1$. Left-handed fermions,
labeled $\psi_{Lj}$ for $j=0,1, \ldots, N$, are understood to
be at sites 0 to $N$, and right-handed ones (labeled correspondingly)
from sites 1 to $N+1$. \label{fig:six}}
\end{figure}

The generalization of the calculation of the chiral currents of
eqn. (\ref{eq:leftcurrent}) to
the $N$-site global moose, illustrated in Fig. \ref{fig:six}, is straightforward.
In this case, the transformation properties of the nonlinear sigma-model
fields are
\begin{align}
\Sigma_1 & \to L \Sigma_1 H^\dagger_1 ~,\\
\Sigma_i & \to H_{i-1} \Sigma_i H^\dagger_i~,\ \ \ i=1,\ldots N\\
\Sigma_{N+1} &\to H_N \Sigma_{N+1} R~.
\end{align}
Defining $\Sigma_j = \exp(2i\pi_j/f_j)$, the corresponding infinitesmal
transformations are
\begin{align}
\frac{2\pi_1}{f_1}& \to \frac{2\pi_1}{f_1} + \alpha_L - h_1~,\\
\frac{2\pi_i}{f_i}& \to \frac{2\pi_i}{f_i} + h_{i-1} - h_i~, \ \ \ i=1,\ldots, N \\
\frac{2\pi_{N+1}}{f_{N+1}}& \to \frac{2\pi_{N+1}}{f_{N+1}} + h_N -\alpha_R~,\\
\end{align}
for infinitesmal $SU(2)_{L,R}$ transformations parameterized by $\alpha_{L,R}$.
Unitary gauge corresponds to imposing the condition,
\begin{equation}
f_{i-1} \pi_{i-1}=f_i \pi_i~, \ \ \ i=1,\ldots, N+1~.
\end{equation}

In solving these equations for $\alpha_{L,R}\neq 0$, it is convenient to define
\begin{align}
\frac{1}{F^2} & = \sum_{j=1}^{N+1} \frac{1}{f^2_j}~,\\
\frac{1}{F^2_\ell} & = \sum_{j=1}^{\ell} \frac{1}{f^2_j}~,\\
\frac{1}{\tilde{F}^2_\ell} & = \sum_{j=\ell+1}^{N+1} \frac{1}{f^2_j}~.
\end{align}
We then find that the transformations imply the fermionic currents
\begin{align}
j^{a\mu}_L  = \bar{\psi}_{L0} \frac{\sigma^a}{2} \gamma^\mu \psi_{L0} & +
\sum_{k=1}^N a_k \left[ \bar{\psi}_{Lk} \frac{\sigma^a}{2} \gamma^\mu \psi_{Lk}+
\bar{\psi}_{Rk} \frac{\sigma^a}{2} \gamma^\mu \psi_{Rk}\right]~,\\
j^{a\mu}_R = \bar{\psi}_{N+1k} \frac{\sigma^a}{2} \gamma^\mu \psi_{R\,N+1} & +
\sum_{k=1}^N b_k \left[ \bar{\psi}_{Lk} \frac{\sigma^a}{2} \gamma^\mu \psi_{Lk}+
\bar{\psi}_{Rk} \frac{\sigma^a}{2} \gamma^\mu \psi_{Rk}\right]~,
\end{align}
with
\begin{align}
a_k & = \frac{F^2}{\tilde{F}^2_k}~,\\
b_k & = \frac{F^2}{F^2_k}~.
\end{align}
Note that $a_k + b_k \equiv 1$, and therefore the vector
currents $j^{a\mu}_V = j^{a\mu}_L + j^{a\mu}_R$ are of canonical
form. 

These results could be used in more complicated Higgsless models,
and suggest that the kinds of corrections which we have found in the 3-site
model occur more generally in models with fermion delocalization.

\section{Ward-Takahashi Identities}
\label{sec:wi}

In this appendix we review the Ward-Takahashi identity that forms the basis of the ``gaugeless"
limit used in our computation of the one-loop correction to the $Z\to b\bar{b}$ amplitude.
We begin by reviewing the standard model result \cite{Barbieri:1992nz,Barbieri:1992dq},
and then present the generalization to the three-site model.

\subsection{Standard Model \cite{Barbieri:1992nz,Barbieri:1992dq}}

In the gaugeless limit of the standard model (SM), the $Z$-boson (as a
classical field) couples to the conserved current
\begin{align}
J^{\mu} = \hat{J}^{\mu}-M_{Z} \partial^{\mu}\pz,
\end{align}
where $\hat{J}^{\mu}$ is the fermionic contribution to the current
\begin{align}
\hat{J}^{\mu}=
 g_{\tbox{Zbb}}^{\tbox{L}} \bLb\gamma^{\mu}\PL\bL
+g_{\tbox{Zbb}}^{\tbox{R}} \bRb\gamma^{\mu}\PR\bR
+\cdots.
\end{align}
The Ward-Takahashi identity arising from the current conservation is
\begin{align}
\partial_{\mu}^x
\vev{T \hat{J}^{\mu}(x)b(y)\overline{b}(z)}
= &
M_Z  \vev{T (\Box_x\pz(x))b(y)\overline{b}(z)}
  -
\delta(x-y)
\left(
g_{\tbox{Zbb}}^{\tbox{L}}\PL + g_{\tbox{Zbb}}^{\tbox{R}}\PR
\right)
\vev{b(x)\overline{b}(z)} \nonumber
\\
& 
\qquad +
\delta(x-z)
\vev{b(y)\overline{b}(x)}\left(
g_{\tbox{Zbb}}^{\tbox{L}}\PR + g_{\tbox{Zbb}}^{\tbox{R}}\PL
\right)~.
\label{eq:Ward-SM-space}
\end{align}
In momentum space, we have the relationship between the (connected, but not 1PI) Green's functions
\begin{align}
\ii (p_1+p_2)_{\mu}\vev{\hat{J}^{\mu}(p_1+p_2)b(p_2)\overline{b}(p_1)}
=
& -M_Z(p_1+p_2)^2
\vev{\pz(p_1+p_2)b(p_2)\overline{b}(p_1)} \nonumber
\\ 
&
-
\left(
g_{\tbox{Zbb}}^{\tbox{L}}\PL + g_{\tbox{Zbb}}^{\tbox{R}}\PR
\right)
S_{bb}(-p_2)
+
S_{bb}(p_1)
\left(
g_{\tbox{Zbb}}^{\tbox{L}}\PR + g_{\tbox{Zbb}}^{\tbox{R}}\PL
\right)~,
\label{eq:Ward-SM-momentum}
\end{align}
where  $S_{bb}(p)$ is the $b$-quark propagator. 
We can also write eqn.~(\ref{eq:Ward-SM-momentum}) in terms
of the 1PI Green's functions
\begin{align}
\ii (p_1+p_2)_{\mu}\vev{\hat{J}^{\mu}(p_1+p_2)b(p_2)\overline{b}(p_1)}_{\tbox{1PI}}
 & = 
-
\ii M_Z
\vev{\pz(p_1+p_2)b(p_2)\overline{b}(p_1)}_{\tbox{1PI}}
\nonumber\\
&
-
S_{bb}^{-1}(p_1)
\left(
g_{\tbox{Zbb}}^{\tbox{L}}\PL + g_{\tbox{Zbb}}^{\tbox{R}}\PR
\right)
+
\left(
g_{\tbox{Zbb}}^{\tbox{L}}\PR + g_{\tbox{Zbb}}^{\tbox{R}}\PL
\right)
S_{bb}^{-1}(-p_2).
\label{eq:Ward-SM-momentum-amp}
\end{align}
We decompose the amputated Green's functions according to their
Lorentz structures\footnote{In  eqn.~(\protect\ref{eq:gamma5structure}) we note that
the amplitude has, in general, both scalar and pseudoscalar parts. In the conventional
bases, the fermion masses are real and the scalar part vanishes.} as
\begin{align}
S^{-1}_{bb}(p)
&=
-\ii\left(\slh{p} A_{bb}(p^2)-B_{bb}(p^2)\right)
\nonumber \\
&\simeq
-\ii \left(\slh{p} A_{bb}(m_b^2)
-B_{bb}(m_b^2)+\mathcal{O}(p^2-m_b^2)\right),
\\
\vev{\hat{J}^{\mu}(p_1+p_2)b(p_2)\overline{b}(p_1)}_{\tbox{1PI}}
&=
\gamma^{\mu}\left(\vev{\hat{J}^{\mu}b\overline{b}}|_{\gamma^{\mu}}\right),
\\
\vev{\pz(p_1+p_2) b(p_2)\overline{b}(p_1)}_{\tbox{1PI}}
&=
(\slh{p}_1+\slh{p}_2)\left(\vev{\pz b\overline{b}}|_{\slh{p}}\right)
+\gamma_5 \left(\left.\vev{\pz b\overline{b}}\right|_{\gamma_5}\right),
\label{eq:gamma5structure}
\end{align}
and where the components of the propagator can be decomposed into two chirality components
\begin{align}
A_{bb}(p^2) &= A_{bb}^{\tbox{L}}(p^2)\PL + A_{bb}^{\tbox{R}}(p^2)\PR,
\nonumber\\
B_{bb}(p^2) &= B_{bb}^{\tbox{L}}(p^2)\PL + B_{bb}^{\tbox{R}}(p^2)\PR.
\end{align}
The Ward-Takahashi identity in eqn.~(\ref{eq:Ward-SM-momentum-amp}) then gives us the conditions
\cite{Barbieri:1992nz,Barbieri:1992dq}
\begin{align}
\ii\vev{\hat{J}^{\mu} b\overline{b}}|_{\gamma^{\mu}} &= -\ii M_Z
\vev{\pz b\overline{b}}|_{\slh{p}} + \ii A_{bb}(m_b^2) \left(
g_{\tbox{Zbb}}^{\tbox{L}}\PL + g_{\tbox{Zbb}}^{\tbox{R}}\PR
\right),
\label{eq:Ward-p1}
\\
0 &= -M_Z  \gamma_5 \vev{\pz b \overline{b}}|_{\gamma_5}
-
B_{bb}(m_b^2) \left(
g_{\tbox{Zbb}}^{\tbox{L}}-g_{\tbox{Zbb}}^{\tbox{R}} \right)
\left(\PL-\PR\right),
\label{eq:Ward-p2}
\end{align}
where we can project out different chiral structures in eqn.~(\ref{eq:Ward-p1}).

At tree level in the standard model we have,
\begin{align}
\vev{\pz b\overline{b}}|_{\slh{p}}^{\tbox{tree}} &= 0,
\\
A_{bb}(0)^{\tbox{tree}} &= 1,
\\
B_{bb}(0)^{\tbox{tree}} &= m_b
\end{align}
so eqn.~(\ref{eq:Ward-p1}) gives us the tree-level $Zb\overline{b}$ coupling,
and
eqn.~(\ref{eq:Ward-p2}) relates the tree-level $\pz b\overline{b}$ coupling
to the mass of the bottom quark.

To compute the $Z\rightarrow\bLb\bL$ amplitude at one-loop order, we must
multiply eqn.~(\ref{eq:Ward-p1}) by the wavefunction renormalization of the
bottom quark according to the LSZ reduction formula and
project out the left-handed chirality component to find
\begin{align}
\ii g_{\tbox{Z$\bLb\bL$}}^{\tbox{1-loop}}\PL
&=
\ii\sqrt{Z_b}\vev{\hat{J}^{\mu} b\overline{b}}|_{\gamma^{\mu}}\sqrt{Z_b}\PL
\nonumber\\
&= -\ii M_Z
\sqrt{Z_b}\vev{\pz b\overline{b}}|_{\slh{p}}\sqrt{Z_b}\PL +
\ii
\sqrt{Z_b}A_{bb}(m_b^2)
\left(
g_{\tbox{Zbb}}^{\tbox{L}}\PL + g_{\tbox{Zbb}}^{\tbox{R}}\PR
\right)
\sqrt{Z_b}\PL.
\label{eq:Ward-loop}
\end{align}
At  one-loop order we may write the wavefunction renormalization as
\begin{align}
\sqrt{Z_b}
&= \sqrt{Z_{\bL}}\PL + \sqrt{Z_{\bR}}\PR
\nonumber\\
&= 1 + \frac{1}{2}\delta Z_{\bL}\PL + \frac{1}{2}\delta Z_{\bR}\PR
\end{align}
where the $\delta Z_{\bL,\bR}$ are of one-loop order.
We also note that, to one-loop order, we have
\begin{align}
A_{bb}(p^2) = 1 - \delta Z_{\bL}\PL - \delta Z_{\bR}\PR.
\end{align}
The $Z\bL\bL$ coupling at one-loop order is then,
\begin{align}
\ii g_{\tbox{Z$\bLb\bL$}}^{\tbox{1-loop}}
=
\ii
g_{\tbox{Zbb}}-\ii M_Z
\vev{\pz b\overline{b}}|_{\slh{p}}.
\label{eq:Ward-loop2}
\end{align}
Note that the corrections due to the wavefunction renormalization
have canceled, and we have only to calculate the $\pz\bLb\bL$
coupling to one-loop \cite{Barbieri:1992nz,Barbieri:1992dq}.

\subsection{Ward-Takahashi Identity in the Three-Site Higgsless Model}
In the gaugeless limit of the three-site model, the $Z$-boson couples to the conserved current
\begin{align}
J^{\mu}_{\tbox{3-site}}
=
{\hat J}^{\mu}_{\tbox{3-site}}
-
M_Z
\partial ^{\mu} \pz,
\end{align}
with  fermionic contributions which now
involve both ``diagonal" and ``off-diagonal" terms
\begin{align}
{\hat J}^{\mu}_{\tbox{3-site}}
&=
\overline{b}
\gamma^{\mu}
g_{\tbox{Zbb}}
b
+
\left(
\overline{B}
\gamma^{\mu}
g_{\tbox{ZBb}}
b
+\mbox{h.c.}
\right)
+\cdots~,
\end{align}
where, for convenience, we define
\begin{align}
g_{\tbox{Zbb}}
\equiv
g_{\tbox{Zbb}}^{\tbox{L}}\PL
+
g_{\tbox{Zbb}}^{\tbox{R}}\PR,
\end{align}
and also the coupling
\begin{align}
\tilde{g}_{\tbox{Zbb}}
\equiv
g_{\tbox{Zbb}}^{\tbox{L}}\PR
+
g_{\tbox{Zbb}}^{\tbox{R}}\PL,
\end{align}
and similarly define $g_{\tbox{ZBb}}$ and $\tilde{g}_{\tbox{ZBb}}$.
The Ward-Takahashi identity
arising from the current conservation is
\begin{align}
\partial_{\mu}^x
\vev{T \hat{J}^{\mu}_{\tbox{3-site}}(x)b(y)\overline{b}(z)}
= &
M_Z  \vev{T (\Box_x\pz(x))b(y)\overline{b}(z)}
-
\delta(x-y)
\left[
g_{\tbox{Zbb}}
\vev{b(x)\overline{b}(z)}
+
g_{\tbox{ZBb}}
\vev{B(x)\overline{b}(z)}
\right]
\nonumber\\
&
+
\delta(x-z)
\left[
\vev{b(y)\overline{b}(x)}
\tilde{g}_{\tbox{Zbb}}
+
\vev{b(y)\overline{B}(x)}
\tilde{g}_{\tbox{ZBb}}
\right]~.
\label{eq:Ward-3S-space}
\end{align}
In momentum space, we have the relationship between the (connected) Green's functions
\begin{align}
\ii (p_1+p_2)_{\mu}\vev{\hat{J}^{\mu}_{\tbox{3-site}}(p_1+p_2)b(p_2)\overline{b}(p_1)}
& =
-M_Z(p_1+p_2)^2
\vev{\pz(p_1+p_2)b(p_2)\overline{b}(p_1)}
\nonumber\\
&
-
g_{\tbox{Zbb}}
S_{bb}(-p_2)
-
g_{\tbox{ZBb}}
S_{Bb}(-p_2)
+
S_{bb}(p_1)
\tilde{g}_{\tbox{Zbb}}
+
S_{bB}(p_1)
\tilde{g}_{\tbox{ZBb}}.
\label{eq:Ward-3S-momentum}
\end{align}
Note that, compared to the SM (see eqn. (\ref{eq:Ward-SM-momentum})), 
eqn. (\ref{eq:Ward-3S-momentum}) contains
additional terms $S_{Bb}$ and $S_{bB}$ because $\hat{J}_{\mu}$ contains
$\bar{B}\gamma^{\mu}b$ and $\bar{b}\gamma^{\mu}B$ contributions.
Also, as in the case of the SM, the Green's functions and couplings
are four-component matrices in Dirac spinor space which can be separated in
terms of their chiral structure.

In the presence of $B-b$ mixing,
we have to take into account that the
connected Green's functions can involve
non-1PI fermion-mixing diagrams .
For example,
\begin{align}
\vev{\hat{J}^{\mu}_{\tbox{3-site}}(p_1+p_2)b(p_2)\overline{b}(p_1)}
&=
S_{bb} \vev{\hat{J}^{\mu}_{\tbox{3-site}}b\overline{b}}_{\tbox{1PI}} S_{bb}
+
S_{bB} \vev{\hat{J}^{\mu}_{\tbox{3-site}}B\overline{b}}_{\tbox{1PI}} S_{bb}
\nonumber\\
&\quad
+
S_{bb} \vev{\hat{J}^{\mu}_{\tbox{3-site}}b\overline{B}}_{\tbox{1PI}} S_{Bb}
+
S_{bB} \vev{\hat{J}^{\mu}_{\tbox{3-site}}B\overline{B}}_{\tbox{1PI}} S_{Bb},
\end{align}
where $\langle\ \rangle_{\tbox{1PI}}$, as before, are the 1PI Green's functions,
and where $S_{bB}(p)$ is the ``off-diagonal" fermion propagator.
The Ward-Takahashi identity involving the 1PI  Green's functions is then
\begin{align}
\ii (p_1+p_2)_{\mu}
&\Big\{
\vev{\hat{J}^{\mu}_{\tbox{3-site}}(p_1+p_2)b(p_2)\overline{b}(p_1)}_{\tbox{1PI}}
\nonumber\\
&\quad
+
S_{bb}^{-1}(p_1)S_{bB}(p_1)\vev{\hat{J}^{\mu}_{\tbox{3-site}}(p_1+p_2)b(p_2)\overline{B}(p_1)}_{\tbox{1PI}}
\nonumber\\
&\quad
+
\vev{\hat{J}_{\tbox{3-site}}(p_1+p_2)B(p_2)\overline{b}(p_1)}_{\tbox{1PI}} S_{Bb}(-p_2) S_{bb}^{-1}(-p_2)
\nonumber\\
&\quad
+
S_{bb}^{-1}(p_1)S_{bB}(p_1)
\vev{\hat{J}^{\mu}_{\tbox{3-site}}(p_1+p_2)B(p_2)\overline{B}(p_1)}_{\tbox{1PI}}
S_{Bb}(-p_2) S_{bb}^{-1}(-p_2)
\Big\}
\nonumber\\
=
-\ii M_Z
&\Big\{
\vev{\pz(p_1+p_2) b(p_2)\overline{b}(p_1)}_{\tbox{1PI}}
+
S_{bb}^{-1}(p_1) S_{bB}(p_1) \vev{\pz(p_1+p_2) b(p_2)\overline{B}(p_1)}_{\tbox{1PI}}
\nonumber\\
&\quad
+
\vev{\pz(p_1+p_2) B(p_2)\overline{b}(p_1)}_{\tbox{1PI}} S_{Bb}(-p_2) S_{bb}^{-1}S_{Bb}(-p_2)
\nonumber\\
&\quad
+
S_{bb}^{-1}(p_1)S_{bB}(p_1)
\vev{\pz(p_1+p_2) B(p_2)\overline{B}(p_1)}_{\tbox{1PI}}
S_{Bb}(-p_2) S_{bb}^{-1}(-p_2)
\Big\}
\nonumber\\
&\quad
- S_{bb}^{-1}(p_1)
g_{\tbox{Zbb}}
- S_{bb}^{-1}(p_1)
g_{\tbox{ZBb}}
S_{Bb}(-p_2)
S_{bb}^{-1}(-p_2)
\nonumber\\
&\quad
+
\tilde{g}_{\tbox{Zbb}}S_{bb}^{-1}(-p_2)
+
S_{bb}^{-1}(p_1)
S_{bB}(p_1)
\tilde{g}_{\tbox{ZBb}}
S_{bb}^{-1}(-p_2)~.
\label{eq:Ward-3S-momentum-amp}
\end{align}

We now work to one-loop order with $p^2_{1,2}\sim m_b^2$.
At this order, we have
\begin{align}
S_{bb}^{-1}(p)S_{bB}(p) &= (-\ii \Sigma_{bB}(p))S^{\tbox{tree}}_{BB}(p)
= \Sigma_{bB}(p)\frac{1}{\slh{p}-M_B},
\\
S_{Bb}(p) S_{bb}^{-1}(p) &= S^{\tbox{tree}}_{BB}(p)(-\ii \Sigma_{Bb}(p))
= \frac{1}{\slh{p}-M_B}\Sigma_{Bb}(p)~,
\end{align}
where $\Sigma_{bB}$ is the fermion-mixing self-energy function and
$S_{BB}(p)$ is the heavy $B$-quark propagator. Let
\begin{align}
\Sigma_{bB}(p) = -\slh{p}\ \delta Z_{bB} + \delta M_{bB},
\end{align}
where, again,  $\delta Z_{bB}$ and $\delta M_{bB}$ are matrices in four-component
Dirac spinor space.
The left-hand side of eqn.~(\ref{eq:Ward-3S-momentum-amp}) simplifies to
\begin{align}
&\quad
\( \mbox{LHS of eqn.~\ref{eq:Ward-3S-momentum-amp}} \)
\nonumber\\
&=
\ii (p_1+p_2)_{\mu}
\left\{
\vev{\hat{J}^{\mu}_{\tbox{3-site}}b\overline{b}}_{\tbox{1PI}}
+
\Sigma_{bB}(p_1)\frac{1}{\slh{p}_1-M_B}\gamma^{\mu} g_{\tbox{ZBb}}
+
\gamma^{\mu} g_{\tbox{ZBb}} \frac{1}{-\slh{p}_2-M_B}\Sigma_{Bb}(-p_2)
\right\}.
\label{eq:Ward-3S-momentum-LHS}
\end{align}
The right-hand side of
eqn.~(\ref{eq:Ward-3S-momentum-amp}) simplifies to
\begin{align}
\(\mbox{RHS of eqn.~\ref{eq:Ward-3S-momentum-amp}}\)
&=
-M_{Z}
\vev{\pz b\overline{b}}_{\tbox{1PI}}
-
\ii S_{bb}^{-1}(p_1)
g_{\tbox{Zbb}}
+
\tilde{g}_{\tbox{Zbb}}
S_{bb}^{-1}(-p_2)
\nonumber\\
&\quad
+
\Sigma_{bB}(p_1)\frac{1}{\slh{p}_1-M_B}
\left\{
\ii M_Z g_{\pz\tbox{$\overline{B}b$}}
-
\tilde{g}_{\tbox{ZBb}}S^{-1}_{bb}(-p_2)
\right\}
\nonumber\\
&\quad
+
\left\{
\ii M_Z g_{\pz\tbox{$\overline{b}B$}}
+
S^{-1}_{bb}(p_1)g_{\tbox{ZBb}}
\right\}
\frac{1}{-\slh{p}_2-M_B}
\Sigma_{Bb}(-p_2)~.
\label{eq:Ward-3S-momentum-RHS1}
\end{align}
Using the tree-level relations in the limit of vanishing $m_b$
\begin{align}
M_{Z}g_{\pz\tbox{$\overline{b}B$}} &= -\ii M_B \tilde{g}_{\tbox{ZBb}},
\\
M_{Z}g_{\pz\tbox{$\overline{B}b$}} &= \ii M_B g_{\tbox{ZBb}},
\\
S_{bb}^{-1}(p) &= -\ii\slh{p},
\end{align}
we can further simplify eqn.~(\ref{eq:Ward-3S-momentum-RHS1})
\begin{align}
&\quad
\begin{pmatrix} \mbox{RHS of eqn.~\ref{eq:Ward-3S-momentum-amp}} \end{pmatrix}
\nonumber\\
&=
-M_{Z}
\vev{\pz b\overline{b}}_{\tbox{1PI}}
-
S_{bb}^{-1}(p_1)
g_{\tbox{Zbb}}
+
\tilde{g}_{\tbox{Zbb}}
S_{bb}^{-1}(-p_2)
\nonumber\\
&\quad
+
\ii\Sigma_{bB}(p_1)
\frac{1}{\slh{p}_1-M_B}
(\slh{p}_1+\slh{p}_2)
g_{\tbox{ZBb}}
+
\ii
(\slh{p}_1+\slh{p}_2)
g_{\tbox{ZBb}}
\frac{1}{-\slh{p}_2-M_B}
\Sigma_{Bb}(-p_2)
\nonumber\\
&
-
\frac{M_Z}{M_B}
\Sigma_{bB}(p_1)
g_{\pz\tbox{$\overline{B}b$}}
-
\frac{M_Z}{M_B}
g_{\pz\tbox{$\overline{b}B$}}
\Sigma_{Bb}(-p_2).
\label{eq:Ward-3S-momentum-RHS2}
\end{align}
Combining Eqs.~(\ref{eq:Ward-3S-momentum-LHS}) and
(\ref{eq:Ward-3S-momentum-RHS2}),
we have three-site Ward-Takahashi identity at one-loop (for $m_b=0$)
\begin{align}
\quad \ii (p_1+p_2)_{\mu}
\vev{\hat{J}^{\mu}_{\tbox{3-site}}b\overline{b}}_{\tbox{1PI}}
&=
-M_{Z}
\Big\{
\vev{\pz b\overline{b}}_{\tbox{1PI}}
+
\frac{\Sigma_{bB}(p_1)}{M_B}
g_{\pz\tbox{$\overline{B}b$}}
+
g_{\pz\tbox{$\overline{b}B$}}
\frac{\Sigma_{Bb}(-p_2)}{M_B}
\Big\}
\nonumber\\
&\quad
-
S_{bb}^{-1}(p_1)
g_{\tbox{Zbb}}
+
\tilde{g}_{\tbox{Zbb}}
S_{bb}^{-1}(-p_2)~.
\label{eq:Ward-3S-momentum-Final}
\end{align}
Compared to the Ward-Takahashi Identity in the standard model
(eqn.~(\ref{eq:Ward-SM-momentum-amp})), we have
the additional terms due to the $B-b$ mixing.
At tree-level, however, these effects vanish and we simply
have the standard model results.
To compute the $Z\bLb\bL$ amplitude at one-loop,
we have to separate the different contributions
of the
amplitude according to their Lorentz structure,
and collect terms proportional to $(\slh{p}_1+\slh{p}_2)$.
Since the leading term in the top line as well as the terms in the bottom line of 
eqn.~(\ref{eq:Ward-3S-momentum-Final})
are present in the standard model Ward-Takahashi identity,
their contribution to the $Z\bLb\bL$ amplitude
is presented in eqn.~(\ref{eq:Ward-loop2}).
(There are,
however,
additional diagrams in the three-site model that contribute to
$\vev{\pz b\overline{b}}$.)
As for the contributions due to $B-b$ mixing,
it is only the kinetic mixing contributions of
$\Sigma_{bB}$ that contribute
to $Z\bLb\bL$ because we are interested only in terms
proportional to $\slh{p}$ in
$\Sigma_{bB}$.
We note that
\begin{align}
\Big\{\Sigma_{Bb}(p)\PL\Big\}\Big|_{\slh{p}} &=
\Big\{\Sigma_{bB}(p)\PL\Big\}\Big|_{\slh{p}} = -\delta Z_{\bL\BL},\\
g_{\pz\tbox{$\overline{B}b$}} &= - \tilde{g}_{\pz\tbox{$\overline{b}B$}},
\end{align}
so the $Z\bL\bL$ coupling of the three-site model at one-loop order is
\begin{align}
\ii g_{\tbox{$Z\bLb\bL$}}^{\tbox{3-Site}}
=
\ii
g_{\tbox{$Z\bLb\bL$}}-\ii M_Z
\Big\{
\vev{\pz b\overline{b}}|_{\slh{p}}
-
g_{\pz\tbox{$\BLb\bL$}}\frac{\delta Z_{\bL\BL}}{M_B}\Big\}.
\label{eq:3S-Ward-loop2}
\end{align}
In terms of Feynman diagrams, this is the same as the calculation
performed in the body of the paper.

\section{Unitary Gauge Calculation}
\label{sec:unitary}

\begin{figure}[h!t]
\begin{center}
\includegraphics [width=2.0in]{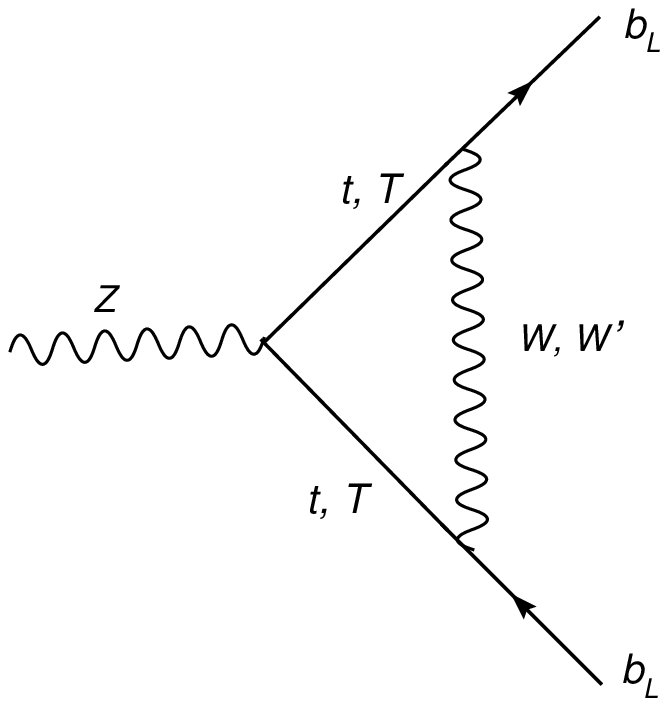}
\includegraphics [width=2.0in]{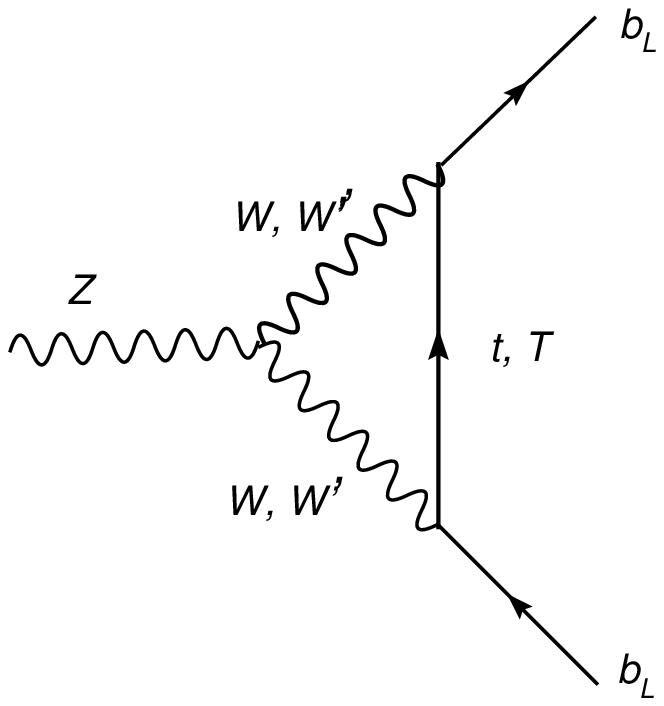}\\
\vspace{0.125in}
\includegraphics [width=4.0in]{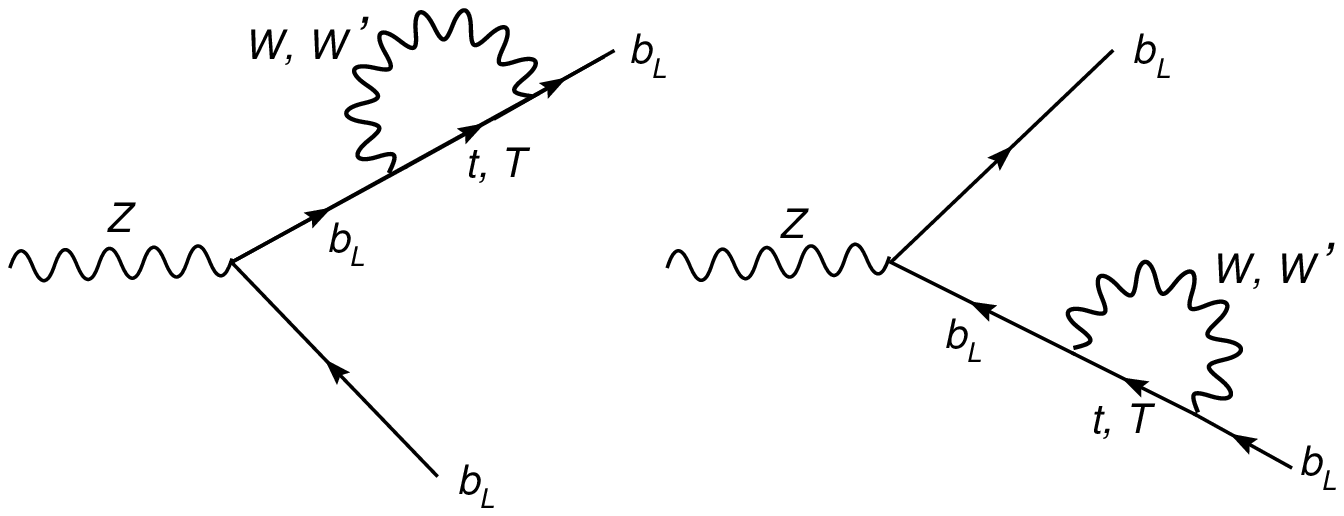}\\
\vspace{0.125in}
\includegraphics [width=4.0in]{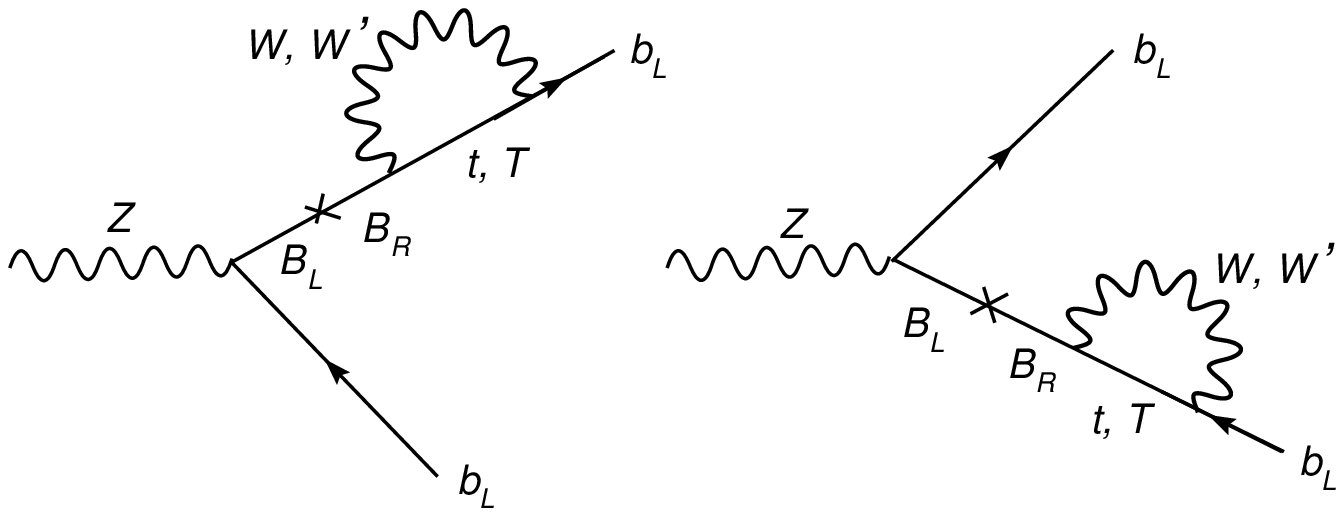}
\caption{
Diagrams that contribute to $\delta g_{Zbb}$
in the unitary gauge. The mass and wavefunction mixing diagrams
implement the results of eqn. (\protect\ref{eq:zbb-couplings}).
}
\label{fig:CTEQ66GGH}
\end{center}
\end{figure}

In this appendix we briefly describe the computation of $Z \to b\bar{b}$
in unitary gauge, without recourse to the gaugeless limit. In order to
do so we must consider the effects of mass and wavefunction mixing
in the bottom sector.
At the one-loop level, the tree-level mass eigenstate $\bL$ mixes with
both $\BL$ and $\BR$, and these mixings give contributions (in
addition to those from the triangle diagrams) to the tree-level
$Z$-boson couplings
\begin{align}
\mathcal{L}_Z
&=g_{\tbox{Zbb}}^{\tbox{L}} Z_{\mu}\bLb\gamma^{\mu}\bL +
g_{\tbox{ZBb}}^{\tbox{L}}Z_{\mu}\left(\bLb\gamma^{\mu}\BL+\BLb\gamma^{\mu}\bL\right).
\end{align}
Let us parameterize the mass and wavefunction mixing through
the Lagrangian\footnote{Here we neglect terms that will not contribute
to the $Z\to b\bar{b}$ process that we are computing. Also, for completeness, we
include contributions of order $m_b$ in the right-handed sector; these are not necessary
for the computation here, but would be necessary to implement the calculation
described in footnote 11.}
\begin{align}
\mathcal{L}
&=
\ii \begin{pmatrix}\bLb \\ \BLb \end{pmatrix}^{T}
\slashi{\partial}
\begin{pmatrix} 1+\delta Z_{\bL} & \delta Z_{\BL\bL} \\ \delta Z_{\BL\bL} & 1 \end{pmatrix}
\begin{pmatrix}\bL \\ \BL \end{pmatrix}
+\ii \begin{pmatrix}\bRb \\ \BRb \end{pmatrix}^{T}
\slashi{\partial}
\begin{pmatrix} 1 & 0 \\ 0 & 1 \end{pmatrix}
\begin{pmatrix} \bR \\ \BR \end{pmatrix} \nonumber
\\
&\quad
-
\left[\begin{pmatrix}\bLb & \BLb \end{pmatrix}
\begin{pmatrix} m_b + \delta m_{b} & \delta M_{bB} \\ 0 & M \end{pmatrix}
\begin{pmatrix}\bR \\ \BR \end{pmatrix}
+\mbox{h.c.}\right].
\label{eq:L1}
\end{align}
The one-loop canonical mass eigenstates (denoted with a superscript $r$) are related
to the tree-level eigenstates by
\begin{align}
\begin{pmatrix}
\bL \\ \BL
\end{pmatrix}
&=
\begin{pmatrix}
1-\frac{\delta Z_{\bL}}{2} \quad & \frac{\delta M_{bB}}{M}-\delta Z_{\BL\bL}
\\
-\frac{\delta M_{bB}}{M}\quad & 1
\end{pmatrix}
\begin{pmatrix}
\bL^{r} \\ \BL^{r}
\end{pmatrix},
\label{eq:bleft}
\\
\begin{pmatrix}
\bR \\ \BR
\end{pmatrix}
&=
\begin{pmatrix}
1 & \frac{m_b}{M}\left(\frac{\delta M_{bB}}{M}-\delta Z_{\BL\bL}\right)
\\
-\frac{m_b}{M}\left(\frac{\delta M_{bB}}{M}-\delta Z_{\BL\bL} \right) & 1
\end{pmatrix}
\begin{pmatrix}
\bR^{r} \\ \BR^{r}
\end{pmatrix}.
\label{eq:bright}
\end{align}

In terms of the one-loop mass eigenstates (eqns. (\ref{eq:bleft}) and (\ref{eq:bright})), 
we find
\begin{align}
\mathcal{L}_Z
&=
\left[
g_{\tbox{Zbb}}^{\tbox{L}}\left(1-\delta Z_{\bL}\right)-
2g_{\tbox{ZBb}}^{\tbox{L}}\frac{\delta M_{bB}}{M}
\right] Z_{\mu}\bLb^{r}\gamma^{\mu}\bL^{r}.
\label{eq:zbb-couplings}
\end{align}
Computing the diagrams illustrated in Fig.~\ref{fig:CTEQ66GGH}
in unitary gauge, using dimensional regularization and $\overline{\rm MS}$,
and subtracting the corresponding diagrams for the $d$ or $s$ quarks
(to isolate the flavor-dependent correction), we reproduce eqn. 
(\ref{eq:result}).

\end{document}